\documentclass[11pt,a4paper]{article}

\usepackage[utf8]{inputenc}
\usepackage[headheight=15pt,left=3cm,right=3cm,top=3cm,bottom=2.5cm]{geometry}
\usepackage{apacite}
\usepackage{amsmath}
\usepackage{amssymb,bm}
\usepackage{amsthm}
\usepackage{natbib}
\usepackage{multirow}
\usepackage{tabularx}
\usepackage{booktabs}
\usepackage{graphicx}
\usepackage{etoolbox}
\usepackage{placeins}
\usepackage{enumerate}
\usepackage{color}
\usepackage{fancyvrb}
\usepackage{bbm}
\usepackage{hyperref}
\usepackage[capitalise]{cleveref}

\usepackage{dcolumn}
\usepackage{siunitx}
\usepackage{mathabx}
\newcolumntype{d}[1]{D{.}{.}{#1}}

\newtheorem{algorithm}{Algorithm}

\renewcommand{\top}{{\mkern-1.5mu\mathsf{T}}}

\newcolumntype{L}[1]{>{\raggedright\let\newline\\\arraybackslash\hspace{0pt}}m{#1}}
\newcolumntype{C}[1]{>{\centering\let\newline\\\arraybackslash\hspace{0pt}}m{#1}}
\newcolumntype{R}[1]{>{\raggedleft\let\newline\\\arraybackslash\hspace{0pt}}m{#1}}

\usepackage{authblk}

\title{Longitudinal analysis of exchanges of support between parents and children in the UK}

\author[1]{Fiona Steele}
\author[2]{Siliang Zhang\footnote{Corresponding author}}
\author[3]{Emily Grundy}
\author[4]{Tania Burchardt}

\affil[1]{Department of Statistics, London School of Economics \& Political Science}
\affil[2]{School of Statistics, KLATASDS-MOE, East China Normal University}
\affil[3]{Institute for Economic and Social Research, University of Essex}
\affil[4]{Centre for Analysis of Social Exclusion, London School of Economics \& Political Science}
\date{}

\begin{document}
\maketitle 

\begin{abstract}
We consider how exchanges of support between parents and adult children vary by demographic and socio-economic characteristics and examine evidence for reciprocity in transfers and substitution between practical and financial support. Using data from the UK Household Longitudinal Study 2011-19, repeated measures of help given and received are analysed jointly using multivariate random effects probit models.  Exchanges are considered from both a child and parent perspective. In the latter case, we propose a novel approach to account for correlation between mother and father reports and develop an efficient MCMC algorithm suitable for large datasets with multiple outcomes.
\end{abstract}

\section{Introduction}
\label{sec:intro}

Population ageing, and the increased difficulties faced by many young adults in accessing secure housing and employment, have prompted a growing debate about intergenerational equity focused largely on public resource transfers \citep{gardiner.etal.2020, willetts.2019}.  However, public resource transfers, which in advanced industrial societies tend to be upward, interact with private transfers within families, which tend to be downward, although with some reversal after around age 75 \citep{kalmijn.2019,lee.2020}, and have large, and possibly increasing, impacts on well-being across the life course \citep{steinback.2012}.  Declining mortality over the past century has resulted in the longer co-survival of adult children and parents, not yet substantially offset by later ages at childbearing, and a change in the balance of age groups within kin groups, as in the population \citep{murphy.etal.2006,murphy.2011}. Increases in longevity, female labour force participation, family disruption, and delayed transitions to adulthood have led to greater needs for family help in providing childcare, support for young adults and care for older people \citep{zigante.etal.2021,henretta.etal.2018,grootegoed.vandijk.2012}. These increasing needs, in combination with reduced state intervention may have important implications for inequalities in well-being across and within generations \citep{dykstra.2018}. This is particularly so because the demographic structure of families by social class, geographical region and ethnicity has become increasingly polarised, to date most strongly documented in the US \citep{seltzer.bianchi.2013,schoon.2015}. Moreover, socio-economic and ethnic disparities in health in the UK, and by implication needs for support, persist or indeed may be increasing \citep{nazroo.2015}.  Little is known about ethnic variations in support exchanges in the UK, although important differences have been identified in other populations, especially the US \citep{swartz.2009}.

For all these reasons there is a pressing need to update and extend our understanding of family transfers between older and younger adults and differences in these by demographic and socio-economic characteristics. In the current context of debate about intergenerational equity and social inequality, understanding the reciprocity and symmetry of exchanges between adult children and parents is particularly needed.  Another issue of importance in the context of restrictions in the availability of social care for disabled older people and delayed transitions to adulthood, is the extent of substitution of financial for practical help. Previous studies, particularly from the US, have shown that parents with higher incomes provide more financial help to adult children and also the reverse: adult children with higher incomes or higher levels of education provide more financial help to older parents \citep{fingerman.etal.2015,attias-donfut.etal.2005}. Some studies suggest substitution effects of money and time assistance; for example, \cite{bonsang.2007} found that better-off adult children who were more engaged in labour market work and lived further from parents provided more financial, but less practical, help to parents.  However other studies find a positive association between transfers of time and money from adult children to parents \citep{deindl.brandt.2011}; recent UK evidence on this topic is sparse.

The aims of this paper are to enhance our understanding of these important family transfers in the UK using large-scale nationally representative longitudinal data and novel statistical methods.  We build on earlier empirical research on intergenerational exchanges in the following ways. First, we use household panel data from the UK Household Longitudinal Study for the period 2011-19 which provides data on the support that respondents give to and receive from non-coresident parents and adult children for a large sample size, permitting a more detailed analysis of ethnic differences in exchanges in the UK than has been previously possible. Second, we study the correlates of exchanges between parents and children from the perspective of both generations, using samples of adult child and parent respondents.  This allows us to consider the effects of characteristics of both children and parents because, as for other nationally representative datasets on kin support, little information is available for respondents' relatives living in other households. Third, we separate financial and practical support and employ a joint modelling approach which makes efficient use of all available information, and allows quantification of concurrent reciprocity (as the residual correlation between giving and receiving help) and of substitution or complementarity between these different forms of support (as the residual correlation between giving/receiving financial and practical support).  Moreover, from longitudinal data, it is possible to distinguish correlations among outcomes at a given year (due to unmeasured time-varying characteristics) and correlations due to time-invariant characteristics such as individual stable traits and family norms. Most previous research has focused on a subset of exchanges between parents and children (e.g. exchanges in one direction only), modelled each exchange separately, or defined a single outcome to capture different kinds of exchange (e.g. in different directions or different forms of support). Apart from contributions to the study of intergenerational exchanges, we make the following more widely applicable methodological contributions to the analysis of household panel data. The analysis of exchanges from a parental perspective raises two particular challenges: (i) mother and father reports of exchanges with children are correlated, with the between-partner correlation larger than the within-individual correlation, and (ii) coresident couples may form and separate over the observation period.  We propose a non-hierarchical three-level random effects model to handle these features of the data.  Finally, we address the considerable computational challenges of fitting multivariate random effects probit models to large datasets with multiple outcomes by developing an efficient MCMC algorithm which we make available as an R package.

The remainder of the paper is organised as follows.  In Section \ref{sec:previous.research} we provide a brief review of previous research on intergenerational exchanges, with a focus on the UK and Europe. This is followed in Section \ref{sec:data} by a description of the data used in our analysis and the choice of covariates.  The details of the multivariate random effects probit models for respondent-parent (two-level) and respondent-child (three-level with time-varying and time-invariant couple effects), and their estimation, are set out in Section \ref{sec:methods} and applied in Section \ref{sec:analysis}.  Concluding remarks and possible directions for future research are given in Section \ref{sec:discussion}.

\section{Literature review and research questions}
\label{sec:previous.research}

Analysis of exchange behaviours in non-coresident family groups requires information both on kin availability (whether a respondent has a parent alive, for example), and on help provided and/or received. Such data sources for nationally representative samples were sparse until late in the last century as censuses and many surveys focused on households rather than kin groups \citep{wolf.1994}. The availability of relevant data sources and research in the area burgeoned in the 21st century resulting in a large number of European and US studies which have considered, and found, reciprocity in parent-child exchanges of help \citep{leopold.raab.2013}. These include studies considering and reporting: socio-economic variations in transfer patterns \citep{attias-donfut.etal.2005,kalmijn.saraceno.2008,fingerman.etal.2012}; impacts of parental divorce on later parent-child relations \citep{kalmijn.2016}; interactions with public policy \citep{dykstra.2018} and, in the US, differences in intergenerational support exchanges by ethnicity \citep{swartz.2009}. However, there is also a large body of research highlighting interactions between public policies, cultural traditions and private transfers which indicates that results from one setting are not always generalisable to another \citep{albertini.etal.2007,brandt.deindl.2013}. To provide context for this study, we therefore largely focus on the relevant UK literature although the methodological developments which are the main focus and contribution of the study are generalisable to other settings.

In the UK nationally representative data on kin availability and support exchanges were collected in several rounds of the British Social Attitudes Survey from the mid-1980s onwards and used in analyses of trends in aspects of intergenerational exchange over time \citep{grundy.shelton.2001}. A specially designed module on kin availability and kin exchange was included in two rounds of the 1999 British Omnibus Survey \citep{grundy.etal.1999} and results showed high levels of provision and receipt of intergenerational help with indications of reciprocity in that those who provided help were more likely to report receiving it and vice versa. Nearly three-quarters of parents aged 50 and over helped their eldest child with domestic tasks, childcare, money, paperwork, shopping or giving lifts, with mothers providing more help with domestic tasks and childcare and fathers more help with paperwork and lifts.  Further analyses \citep{grundy.shelton.2001, grundy.murphy.2006} showed that provision of help to mothers by adult children was positively associated with the respondent having a child under 16, with older age of mother, and very strongly with receipt of help from mother and with proximity (travel time to parent).  When proximity was not controlled, education was also significant with those in the lowest educational group being twice as likely as those in the highest to provide regular help. Odds of providing help to a father were 40\% or more lower where the father's partnership history was disrupted and also lower where both had disrupted partnership histories. There were some variations in children's provision of help by number of siblings suggesting a greater involvement by only children. Although this data set included responses from both parents and children (not in the same families), the relatively small size ($n$=1,800) and cross-sectional design limited scope for much further analysis. In a more recent study based on the larger National Child Development Study (NCDS, 1958 birth cohort), \cite{evandrou.etal.2018} were able to take a longer perspective on reciprocity by examining whether provision of help to parents by individuals aged 50 varied according to whether they reported having received help from their parents between completion of education and age 42. Results showed that a high proportion of respondents had received help from parents with accommodation, money, or childcare (of respondents' children) and that receipt of such help was positively associated with provision of help to parents.

Other UK studies have used datasets of older people to analyse variations in provision and receipt of support to/from children, and in some cases grandchildren, from the perspective of the parent generation. \cite{grundy.2005} examined exchanges of support between parents aged 55-75 and their adult children using data from the Retirement and Retirement Plans Survey. Results showed that between two thirds and three quarters of parents in this age group were involved in some sort of exchange relationship with at least one of their children. Generally, more Third Age parents were providers than recipients of help, but there was a strong reciprocal element to intergenerational exchange with, for example, married parents who provided support to at least one child being twice as likely as those who did not to receive support from a child, after allowance for a range of relevant parental and child characteristics. Parental characteristics associated with a higher probability of providing help included higher income, home ownership and being married or widowed rather than divorced. Comparative work also using the US Health and Retirement Study (HRS) \citep{henretta.etal.2002,grundy.henretta.2006} found that among married parents provision of financial help to children was positively associated with higher socio-economic status (of the parent), that poorer parental health was associated with provision of less practical help, particularly among unmarried parents. In both Britain and the USA women in later midlife who provided help to their adult children were more likely to also help their parents, and vice versa, indicating the importance of family cultures and norms on patterns of intergenerational exchange; a similar positive association between transfers upward and downward has been reported in a more recent UK study based on analysis of the NCDS \citep{vlachantoni.etal.2020}. In other studies, \cite{grundy.read.2012} used data from the English Longitudinal Study of Ageing (ELSA, a sister study to the HRS) to examine variations in parental receipt of support from adult children. Results showed that among fathers aged 60 and over receipt of help from a child was positively associated with lower wealth, being unmarried and long-term illness. Among mothers, having more children and having a daughter were additionally important. Analyses by \cite{ermisch.2014} of data from the British Household Panel Study (BHPS), UK Household Longitudinal Study (UKHLS) and ELSA also indicated that children's provision of help appeared responsive to parental need.

In a more recent analysis of BHPS and UKHLS data from 2001-15, \cite{steele.grundy.2021} studied differences in adult children's reports of help given to, and received from, parents by the child's employment and partnership status, and recent transitions in these, and presence and age of their own children. Results indicated that children with a higher propensity to give help to parents tended to also have a higher propensity to receive help.  Adult children who had experienced recent partnership breakdown, and those with a young child of their own, were also more likely to report receipt of help from parents, although these factors had little influence on provision of help to parents.

To summarise, these previous studies suggest relatively strong reciprocity between giving and receiving support intergenerationally, although limitations in the data on which some of these studies are based mean the evidence is somewhat patchy. The giving and receiving may take place contemporaneously or at different times. We use the term `reciprocity' to refer to exchanges where those who provide help tend to receive help; we use `reciprocity' interchangeably with `symmetry' to emphasise that a positive correlation between giving and receiving help also suggests that individuals who do not provide help tend not to receive help. Our use of these terms is not intended to imply any latent motive in exchanges, for example that giving is conditional on receipt or sets up an expectation of a return.

Existing evidence also indicates that a range of demographic and socio-economic characteristics affect the provision and receipt of support. Specifically, it suggests the provision and receipt of support is influenced by the \emph{needs} of the recipient (for example, older age, ill health, lower income/lack of employment, not having a partner and, among younger recipients, having a young child), the \emph{capacities} of the donor to provide help (for example, geographical proximity, higher income/being employed, home ownership, and being in good health) and also by cultural, family and \emph{social norms}, including norms associated with gender, ethnic and socio-economic position (indicated by level of education), and with sharing responsibilities among siblings.

Using this framework, we address four main questions about the pattern of financial and practical intergenerational exchange in the contemporary UK:
\begin{enumerate}[(1)]
\item How does the provision of help to parents by adult children vary according to demographic and socio-economic characteristics associated with the capacity of the child to provide help, parental need and social norms?
\item How does parental provision of help to children vary by characteristics associated with parents' capacity to help, child need and social norms?
\item Is there evidence of reciprocity in transfers from both the younger and the older generation, i.e. are those who provide help more likely to report receipt of help? 
\item Is there evidence of a substitution effect, i.e. do donors who provide financial assistance provide less practical help, and vice versa?
\end{enumerate}

We additionally consider the effects of geographical proximity in all analyses and whether or not its inclusion influences other associations.

\section{Data}
\label{sec:data}

\subsection{Data on intergenerational exchanges from the UK Household Longitudinal Study}
\label{sec:data.UKHLS}

We use household panel data from the UK Household Longitudinal Study (UKHLS), also known as Understanding Society \citep{UKHLS20}. The survey began in 2009-10 with a sample of the members of approximately 40,000 households. All members of the wave 1 households and their offspring constitute the core sample who are followed wherever they move within the UK.  All household members aged 16 and over are invited to complete the adult survey.  The survey fieldwork period is 24 months, but individual sample members are interviewed at approximately 12-month intervals. Interviews are conducted face-to-face in respondents' homes or through a self-completion online survey.  Information on exchanges of help with relatives living outside a respondent's household was collected as part of the rotating `family network' module which was administered biennially in 2011-13, 2013-15, 2015-17 and 2017-19 (waves 3, 5, 7 and 9).  We use data from all four available waves.

Respondents with at least one non-coresident parent were asked whether they `nowadays' gave the following eight forms of help to their parent(s) `regularly or frequently' using a binary (yes/no) scale: lifts in a car; help with shopping; providing or cooking meals; help with basic personal needs; washing, ironing or cleaning; personal affairs such as paying bills or writing letters; decorating, gardening or house repairs; and financial help.  Where a respondent had both biological and step or adoptive parents alive, the respondents were asked to report on the ones that they had most contact with. Respondents with a non-coresident adult child were asked the same set of questions about help they had received from their child(ren). The same questions were asked about receipt of support from parents, and support given to children, but with `personal needs' replaced by `help with childcare'.  These questions have been used in earlier UK studies, such as the Retirement and Retirement Plans Survey \citep{disney.etal.1997}, the 1999 Omnibus Kin Study \citep{grundy.etal.1999}, and the BHPS \citep{ermisch.2014}. 
%\citep[e.g.][]{henretta.etal.2002,grundy.2005,grundy.etal.1999,nolan.scott.2006} omitted as did not use BHPS.
%omitted Nolan and Scott (2006).

We define a total of eight binary responses from the above questions: four indicators of support that adult child respondents give to and receive from their parent(s), and four indicators of support that parent respondents give to and receive from their adult child(ren). For each generation of respondent and direction of exchange, financial help is measured by a single binary item, while practical help is coded 1 for an exchange of any of the other seven types of support and 0 otherwise. Based on the analysis sample for child respondents (see Section \ref{sec:data.samples}), the proportion of person-wave observations where the respondent reports giving and receiving support regularly or frequently to or from their parents are: 6.1\% for giving financial help, 13.7\% for receiving financial help, 43.4\% for giving practical help, and 35.9\% for receiving practical help.  The corresponding proportions based on parent reports of exchanges with their children are: 29.3\%, 2.5\%, 52.6\% and 36.8\%.  This is consistent to some extent with previous UK and European studies which report that adult children are more likely to receive than provide financial help to parents \citep{albertini.etal.2007}.

In common with other large-scale general-purpose surveys with information on intergenerational exchanges, the UKHLS family networks module has several limitations.  First, child respondents report on exchanges with both parents together, and parent respondents report on exchanges with all children collectively, so we cannot examine differences in exchanges between mothers and fathers and children or between parents and specific children. As far as parents are concerned, it is only a minority that are separated, so decisions about giving are likely to be at least somewhat joint, and receipt of help is likely to benefit both; the lack of specificity about which parent of the respondent they are referring to therefore matters less. The same argument cannot be made for children of the respondent, who are likely to be living independently from one another. Overall, however, we maintain that these data nevertheless provide useful information about reciprocity between the two generations (in a collective sense) within a family.  

A second limitation is that respondents' non-coresident relatives were not interviewed, and little information was collected about them, so we have to rely on reports of help received and given from only one side of the exchange relationship which may be subject to reporting biases; for example, previous research has found that respondents tend to under-report the help they receive and over-report the help given \citep{shapiro.2004,kim.etal.2011}.  We mitigate for this to some extent by studying exchanges of support from both a parental and child perspective using parent and child samples which allow us to consider the effects of parent and child characteristics on exchanges, albeit not simultaneously.  Although there are longitudinal studies which collect data from both parents and children in the same family, such as the German PAIRFAM study \citep{huinink.etal.2011}, the Californian Longitudinal Study of Generations \citep{bengtson.2001} and a number of others \citep{suitor.etal.2017}, there is no equivalent population representative UK source. Another limitation is that data were collected only on parent-child exchanges, so we are unable to study or allow for potentially important exchanges with other family members, for example grandparents, grandchildren and siblings. %Moreover, the UKHLS has a very large sample size and good rates of study retention.

\subsection{Choice of covariates}
\label{sec:data.covariates}

We consider as covariates a range of individual and household demographic and socio-economic characteristics that aim to capture an individual's capacity to give help to their relatives and their potential need for support. As discussed above, previous studies of intergenerational exchanges in the UK, and elsewhere, have reported differences by gender, partnership status, whether the adult child has children themselves, health status, socio-economic status and geographical proximity, as well as, for parents, the number of children they have, and for adult children, the number of siblings they have. We therefore included indicators of all these variables in the analysis. We additionally included indicators of ethnicity. UK research on ethnic differences in intergenerational exchanges based on nationally representative samples is very limited partly due to the small sample sizes used in many earlier studies, but some studies have shown differences between ethnic groups in proximity of adult children to their parents \citep{chan.ermisch.2015}.  Moreover, studies from the USA and Europe have drawn attention to the influence of origins or connections to collectivist societies which may influence resource exchanges among immigrant and minority ethnic groups \citep{wiemers.bianchi.2015} and reported differences in intergenerational exchanges by ethnic origin \citep{schans.komter.2010, bordone.devalk.2016}.

As little information was collected on non-coresident relatives, most variables refer to the survey respondent. The following respondent characteristics were included as predictors of all types of exchange: gender, age, ethnicity (categorised as Asian and Asian British, Black and Black British, Other, or White), whether they have a coresident partner, whether they have a long-term illness that limits their daily activities, employment status (classified as employed or non-employed, the latter including unemployed and economically inactive), highest education level (secondary school only versus post-school qualifications), household tenure (home-owner or social/private renter), and household income (equivalised, adjusted for inflation using the annual Consumer Price Index for the year of interview, and log transformed). The analysis of respondent exchanges with their parent(s) additionally included indicators of the presence and age of the respondent's youngest biological or adopted children, the number of siblings, the age of the oldest parent, whether either parent lives alone and the travel time to the nearest parent.  The analysis of respondent exchanges with their adult child(ren) included the following variables in addition to the basic set: the number of non-coresident adult children, whether the respondent had a surviving parent, and the travel time to the child that the respondent has most contact with.  Finally, dummy variables for wave were included to capture broad time trends. With the exception of gender and ethnicity which were collected only once when a panel member first entered the study, all other variables were measured at each wave and we use their wave-specific values in the analysis.  Descriptive statistics for all covariates are given in Table \ref{tab:descriptives}.

While the geographical proximity (travel time) between parents and children is treated as a covariate in our analysis, we acknowledge that it is likely to be endogenous with respect to exchanges of support: individuals could move closer to their parents or children to facilitate exchanges in either direction, and the (unobserved) decision on where to live may depend on factors that also influence exchanges.  For this reason, we consider models with and without geographical proximity and in Section \ref{sec:analysis.predictors} we comment on changes in the effects of the other covariates when proximity is excluded.

\subsection{The analysis samples}
\label{sec:data.samples}

For the model of respondent-parent(s) exchanges, the analysis sample was first restricted to 82,823 person-wave observations contributed by respondents aged 18 or older who had at least one non-coresident parent.  There were 3509 records from respondents with a non-coresident parent who were living with their other parent; these were mainly younger respondents who had not yet left the parental home, and were excluded because the nature of their exchanges with the non-coresident parent are likely to differ from those of respondents who do not live with a parent.  For similar reasons, a further 8932 records were excluded when the nearest parent lived abroad. Of the remaining 69,050 records, a further 3438 (5\%) were omitted due to missing values on at least one covariate to give a final sample of 65,612 person-wave records contributed by 26,586 individuals. The covariate with the largest amount of missing data ($n$=1156) was the indicator of whether either of the respondent's parents was living alone, which was due to this question only being asked about biological parents while questions about exchanges could refer to any parent figure. Other variables had 0-1\% of missing values.

For the model of respondent-child(ren) exchanges, the initial sample contained 61,427 person-wave records, which was reduced to 59,324 after excluding records where the child whom the respondent had most contact with lived abroad. The final sample, after excluding records with missing covariates (3\%), contains 57,562 person-wave records from 22,456 individuals.

We note that for exchanges with both parents and children, respondents may move in and out of the target population over time.  For example, a respondent becomes eligible for inclusion in the respondent-parent sample after leaving the parental home and becomes ineligible if they later return or following the death of both parents or a parent moving into their household. There are 15,652 person-wave records from 7566 individuals where the respondent has at least one non-coresident parent and child, and therefore appears in both samples.

\begin{table}
\caption{\label{tab:descriptives} Descriptive statistics for explanatory variables: percentage of person-waves or mean (and standard deviation) in samples for analysis of exchanges with non-coresident parents from an adult child perspective and of exchanges with non-coresident adult children from a parent perspective}
%\caption{\label{tab:descriptives} Descriptive statistics}
\begin{footnotesize}
\centering
\fbox{%
\begin{tabular}{lrr}
\hline
Variable        & \multicolumn{1}{c}{Respondent (child)-} & \multicolumn{1}{c}{Respondent (parent)-} \\
                & \multicolumn{1}{c}{parent analysis}      & \multicolumn{1}{c}{child analysis}  \\
\hline
\textbf{Respondent characteristics}             &               & \\
Age (years)                                     & 42.4 (11.3)   & 64.4 (11.6) \\
Female                                          & 57.9          & 57.5 \\
Ethnicity                                       &               & \\
~~~Asian or Asian British                       & 5.3           & 3.3 \\
~~~Black or Black British                       & 2.4           & 2.3 \\
~~~Other                                        & 1.9           & 1.2 \\
~~~White                                        & 90.4          & 93.2 \\
Coresident partner                              & 76.5          & 70.2 \\
Long-term limiting illness                      & 13.1          & 31.9 \\
Post-school education                           & 45.2          & 28.9 \\
Unemployed or economically inactive             & 23.1          & 62.9 \\
Log equivalised annual household income         & 9.84 (0.76)   & 9.81 (0.71) \\
Home owner (vs social/private rent or other)    & 69.8          & 77.5 \\
Child coresidence status                        &               & \\
~~~No children                                  & 24.7          & -- \\
~~~Coresident only                              & 47.8          & -- \\
~~~Coresident and non-coresident                & 12.0          & -- \\
~~~Non-coresident only                          & 15.4          & -- \\
Age of youngest coresident child                &               & \\
~~~No coresident children                       & 40.1          & -- \\
~~~$<$ 2 years                                  & 9.2           & -- \\
~~~2-4 years                                    & 11.2          & -- \\
~~~5-10 years                                   & 15.4          & -- \\
~~~11-16 years                                  & 12.1          & -- \\
~~~$>$ 16 years                                 & 12.0          & -- \\
Any coresident children                         & --            & 24.3 \\
Number of non-coresident children               &               & \\
~~~1                                            & --            & 24.5 \\
~~~2                                            & --            & 43.1 \\
~~~3+                                            & --            & 32.4 \\
Number of siblings                              &               & \\
~~~None                                         & 9.3           & -- \\
~~~1                                            & 34.3          & -- \\
~~~2+                                           & 56.4          & -- \\
\textbf{Parent characteristics}                 &               & \\
Age of oldest parent (years)                    & 70.8 (11.4)   & -- \\
At least one parent lives alone                 & 36.5          & -- \\
Has a surviving parent                          & --            & 30.6 \\
\textbf{Parent-child characteristics}           &               & \\
Travel time to parent/child                     &               & \\
~~~$<$ 15 minutes                               & 41.4          & 43.3 \\
~~~15-30 minutes                                & 19.8          & 21.3 \\
~~~30-60 minutes                                & 11.6          & 11.6 \\
~~~1-2 hours                                    & 9.9           & 9.6 \\
~~~$>$ 2 hours                                  & 17.3          & 14.2 \\
Number of person-waves (individuals)            & 65,612 (26,586)   & 57,562 (22,456) \\
\hline
\end{tabular}}
\end{footnotesize}
\end{table}

\section{Statistical methods}
\label{sec:methods}

\subsection{Previous approaches}
\label{sec:methods.previous}

%We are not aware of any research that has jointly modelled bidirectional exchanges between children and parents from the perspective of both generations.
Most previous research that has studied reciprocity of exchanges has taken one of three approaches: (i) defined a joint outcome for giving and receiving help, often obtained from latent class analysis (e.g. with categories for low and high exchangers or unidirectional support), (ii) modelled the difference between support given and support received \citep{kalmijn.2019,mudrazija.2016}, or (iii) included exchanges in one direction as a predictor of exchanges in the other direction.  Examples of (i) include \cite{hogan.etal.1993} and \cite{silverstein.bengtson.1997}.  This approach models reciprocity directly, but does not permit analysis of the effects of individual characteristics on exchanges in each direction.  Examples of (iii) include \cite{cheng.etal.2015} and \cite{grundy.2005} who look at contemporaneous reciprocity and \cite{silverstein.etal.2002} and \cite{evandrou.etal.2018} who consider reciprocity over the lifecourse by including predictors that measure exchanges in the other direction defined at an earlier time point. The problem with this approach, especially when the predictor and the outcome are defined at the same time, is that the predictor is likely to share unmeasured influences with the outcome, for example family characteristics, leading to correlation between the predictor and the residual term. More recently, \cite{steele.grundy.2021} proposed a joint model for bidirectional exchanges using panel data from BHPS sample members.  However, the focus of that study was to account for unequal spacing of response data in order to estimate the effects of partnership and employment transitions in the previous year, and the analysis was simplified in other respects: exchanges were viewed only from a child perspective, practical and financial support were combined in a single outcome, the sample size was much smaller than for UKHLS, and a limited set of covariates was considered.

We are not aware of any research that has studied bidirectional exchanges between children and parents and considered the perspectives of both generations. We build on previous research by jointly modelling support given by and received from a family member using longitudinal data on exchanges, and distinguishing financial and practical support.  A major advantage of a joint modelling approach is that the estimated correlations between the support outcomes can be used to answer questions about the degree of reciprocity between parents and children, and whether financial and practical support tend to be given or received together or whether one acts as a substitute for the other. Although data were collected from only one generation in a dyad, we consider both a parental and child perspective by modelling exchanges between adult child respondents and their non-coresident parents and between parent respondents and their non-coresident adult children.  Multivariate random effects probit models are used to allow for within-person correlation in exchange outcomes over time. In the analysis of respondent-child exchanges, we propose a flexible random effects model that additionally allows for within-couple correlation in partners' reports of exchanges with their children.

In the following sections we describe multivariate random effects models for exchanges from the perspectives of adult children (Section \ref{sec:methods.RPmodel}) and parents (Section \ref{sec:methods.RCmodel}) and then provide an overview of their estimation (Section \ref{sec:methods.estimation}).

\subsection{Model for respondent-parent exchanges}
\label{sec:methods.RPmodel}

A two-level multivariate random effects response model is used to analyse bidirectional exchanges of financial and practical help between survey respondents and their parent(s). A $C$ superscript is used to indicate variables and parameters in models where the adult child is the respondent. Denote by $y_{rti}^{(C)*}$ a continuous latent variable underlying the observed binary outcome $(y_{rti}^{(C)})$ for outcome $r$ $(r=1,2,3,4)$ at wave $t$ $(t=1,\ldots,T_i^{(C)})$ for individual $i$ $(i=1,\ldots,n^{(C)})$ where $T_i^{(C)}$ is the number of person-waves at which individual $i$ was observed to have a non-coresident parent and $n^{(C)}$ is the number of individuals.  The outcome index $r$ is coded: 1 for giving practical help (GP), 2 for giving financial help (GF), 3 for receiving practical help (RP) and 4 for receiving financial help (RF). The model takes the form

\begin{equation}
\label{eq:RPmodel}
y_{rti}^{(C)*} = \boldsymbol\beta_r^{(C)} \boldsymbol x_{ti}^{(C)} + u_{ri}^{(C)} + e_{rti}^{(C)},
\end{equation}
where $\boldsymbol x_{ti}^{(C)}$ is a column vector of covariates (common to all outcomes) with a row vector of coefficients $\boldsymbol\beta_r^{(C)}$, $u_{ri}^{(C)}$ is an outcome and individual specific random effect, and $e_{rti}^{(C)}$ is a time-varying residual.  Denote by $\boldsymbol u_i^{(C)}$ and $ \boldsymbol e_{ti}^{(C)}$ the vectors formed by respectively stacking $u_{ri}^{(C)}$ and $e_{rti}^{(C)}$ for $r=1,2,3,4$. We assume that each vector follows a multivariate normal distribution: $\boldsymbol u_i^{(C)} \sim N(\bm 0,\bm \Sigma_u^{(C)})$ and $\boldsymbol e_{ti}^{(C)} \sim N(\bm 0, \bm \Sigma_e^{(C)})$.  Denote by $\sigma_{zrr^\prime}^{(C)}$ the elements of the variance-covariance matrix $\bm \Sigma_z^{(C)}$ where $z \in \{u,e\}$, and $\rho_{zrr^\prime}^{(C)}$ $(r \neq r^\prime)$ the correlations.  For identification we set the variances $\sigma_{err}^{(C)}=1$ for all $r$, so that $\sigma_{err^\prime}^{(C)} = \rho_{err^\prime}^{(C)}$ for $r \neq r^\prime$.

The correlations $\rho_{ur r^\prime}^{(C)}$ measure the associations among unmeasured time-invariant influences on $y_{rti}^{(C)*}$ and $y_{r^\prime ti}^{(C)*}$ after adjusting for the effects of $\boldsymbol x_{ti}^{(C)}$, where these omitted variables may be characteristics of the individual $i$, their parent(s) or the family unit.  Four of these correlations provide measures of the extent of reciprocity in exchanges between children and their parents.  For example, $\rho_{u24}^{(C)}$ is the correlation between giving and receiving financial support which is a measure of reciprocity in financial help, while $\rho_{u13}^{(C)}$  measures reciprocity in practical help.  The correlations $\rho_{u14}^{(C)}$ and $\rho_{u23}^{(C)}$ measure the extent to which giving one form of help is reciprocated with the other form of help.  The remaining correlations measure the associations between giving practical and giving financial help $(\rho_{u12}^{(C)})$ and receiving practical and receiving financial help $(\rho_{u34}^{(C)})$, which would be negative if one form of help tends to be substituted with the other.  The correlations $\rho_{er r^\prime}^{(C)}$ measure the associations among unmeasured time-varying influences, or contemporaneous reciprocity of exchanges and substitution of financial help for practical help.

\subsection{Model for respondent-child exchanges}
\label{sec:methods.RCmodel}

There are two methodological challenges that are particular to the analysis of respondent-child exchanges.  These result from the presence of couples in the sample because all adult members of a household are eligible sample members.  The first challenge is that couples are not stable entities over time because individuals may form and dissolve coresidential unions over the observation period.  The second issue is that in most cases we expect married or cohabiting spouses to report on exchanges with the same children and support may be given and received jointly as a couple, leading to high positive within-couple correlations.  A potential solution to the second problem is to define couple-level outcomes that indicate whether either partner gives or receives support.  However, this is only possible where both partners participate in the survey; in our sample, 38\% of respondents at a given wave have a non-responding partner and thus incomplete couple information.  Even when both members of a couple are present, aggregating to the couple level is wasteful and would complicate the inclusion of individual-level covariates.  For these reasons, we define individual-level responses and allow for `couple' effects in our model.

To address the first concern, we define `couple' clusters following the approach of \cite{steele.etal.2019} who proposed the concept of a `super-household' to model within-household associations in longitudinal data.  In the special case of couples, this involves grouping together all observations from individuals who are linked through coresidential unions during the observation period.  Suppose, for example, that individual A was married to individual B at waves 1 and 2, unpartnered at wave 3 and cohabiting with individual C at wave 4.  The observations for (A,B,C) over the four waves would form a couple cluster.   An individual who remains unpartnered throughout the observation period forms their own couple cluster. The 22,456 individuals in the analysis sample are nested within 15,934 couple clusters, 99.9\% of which contain one or two individuals.

After creating couple clusters, we allow for between-partner correlation by specifying a 3-level extension of model (\ref{eq:RPmodel}) with time-invariant and time-varying couple random effects.  A $j$ subscript is added to index couples and a $P$ superscript indicates that parents are now the respondents. Denote by $y_{rtij}^{(P)*}$ the latent variable underlying binary outcome $r$ at wave $t$ $(t=1,\ldots,T_{ij}^{(P)})$ for individual $i$ $(i=1,\ldots,n_j^{(P)})$ in couple cluster $j$ $(j=1,\ldots,n^{(P)})$ where $T_{ij}^{(P)}$ is the number of person-waves at which individual $i$ in couple cluster $j$ was observed to have a non-coresident child, $n_j^{(P)}$ is the number of individuals in couple cluster $j$ (usually 1 or 2) and $n^{(P)}$ is the number of couple clusters.  We specify a multivariate random effects probit model of the form
\begin{equation}
\label{eq:RCmodel}
y_{rtij}^{(P)*} = \boldsymbol\beta_r^{(P)} \boldsymbol x_{tij}^{(P)} + u_{rij}^{(P)} + v_{rj}^{(P)} + w_{rtj}^{(P)}+ e_{rtij}^{(P)},
\end{equation}
where $\boldsymbol x_{tij}^{(P)}$ is a vector of covariates with coefficients $\boldsymbol\beta_r^{(P)}$ for outcome $r$,  $u_{rij}^{(P)}$ is an individual-specific random effect, $v_{rj}^{(P)}$ is a time-invariant couple-level random effect, $w_{rtj}^{(P)}$ is a time-varying couple-level effect and $e_{rtij}^{(P)}$ is a time-varying residual.  As for the respondent-parent(s) model of (\ref{eq:RPmodel}), we stack the random effects for each response to form four vectors $\boldsymbol u_{ij}^{(P)}$,
$\boldsymbol v_{j}^{(P)}$, $\boldsymbol w_{tj}^{(P)}$ and $\boldsymbol e_{tij}^{(P)}$ which are each assumed to follow a 4-dimensional multivariate normal distribution with covariance matrices $\bm \Sigma_u^{(P)}$, $\bm \Sigma_v^{(P)}$, $\bm \Sigma_w^{(P)}$ and $\bm \Sigma_e^{(P)}$.

Model (\ref{eq:RCmodel}) without $w_{rtj}^{(P)}$ would be a multivariate 3-level hierarchical random intercept model, the univariate form of which is commonly used to analyse 3-level longitudinal data.  However, the addition of $w_{rtj}^{(P)}$ is important to avoid an unrealistic covariance structure.  To illustrate this point, consider latent outcomes $y_{rtij}^{(P)*}$ for a couple $(i=1,2)$ observed at two waves $(t=1,2)$. Adjusting for covariate effects, the variances and covariances implied by (\ref{eq:RCmodel}) are shown in Table \ref{tab:cov.structure}. The within-individual (between-wave) covariance is $\sigma_{vrr}^{(P)} + \sigma_{urr}^{(P)}$, the between-partner covariance at a given wave $t$ is $\sigma_{vrr}^{(P)} + \sigma_{wrr}^{(P)}$, and the between-partner covariance for observations at different waves is $\sigma_{vrr}^{(P)}$.  The omission of $w_{rtj}^{(P)}$ would imply $\sigma_{wrr}^{(P)}=0$ which would impose two unrealistic restrictions on the covariances due to the fact that they are composed of non-negative variance parameters: (i) the within-individual covariance is greater than the between-partner covariance, and (ii) the between-partner covariance is the same regardless of whether the outcomes are at the same wave or different waves.  The covariance structure with $w_{rtj}^{(P)}$ is more reasonable in our application because we expect the between-partner covariance between outcomes at the same wave to be high due to the typical scenario where each partner reports on exchanges with the same child(ren).  Empirically, we find that excluding $w_{rtj}^{(P)}$ leads to a tiny estimate for $\sigma_{urr}^{(P)}$ due to restriction (i) as the fitted model attempts (and fails) to capture the covariance structure in the data.
%To illustrate this point, consider latent outcomes $y_{rtij}^*$ for a couple $(i=1,2)$ observed at two waves $(t=1,2)$, omitting the $P$ subscript for simplicity, and denote by $\sigma_z^2$ the random effect variances where $z \in (u,v,w)$. The residual variances and covariances, $\mbox{cov}(y_{tij}^*,y_{t^\prime i^\prime j}^* | \boldsymbol x_{tij},\boldsymbol x_{t^\prime i^\prime j})$, implied by (\ref{eq:RCmodel}) are shown in Table \ref{tab:cov.structure}. The within-individual (between-wave) covariance is $\sigma_v^2 + \sigma_u^2$, the between-partner covariance at a given wave $t$ is $\sigma_v^2 + \sigma_w^2$, and the between-partner covariance for observations at different waves is $\sigma_v^2$.
\begin{table}
\caption{\label{tab:cov.structure} Within-individual and within-couple residual covariance structure for outcome $r$ implied by respondent-child model (\ref{eq:RCmodel}) for a mother-father couple (individuals $i=1, 2$) at waves $t=1, 2$. The total residual variance is denoted by $\sigma_{rr}^{(P)} = \sigma_{vrr}^{(P)} + \sigma_{urr}^{(P)} + \sigma_{wrr}^{(P)} + 1$.}
\begin{footnotesize}
\begin{center}
\fbox{%
\begin{tabular}{lccccc}
\hline
& & \multicolumn{2}{c}{$i=1$} & \multicolumn{2}{c}{$i=2$} \\
\cmidrule(lr){3-4} \cmidrule(lr){5-6}
& & $t=1$ & $t=2$ & $t=1$ & $t=2$ \\
\hline
$i=1$ & $t=1$ & $\sigma_{rr}^{(P)}$ & & & \\
    & $t=2$ & $\sigma_{vrr}^{(P)} + \sigma_{urr}^{(P)}$ & $\sigma_{rr}^{(P)}$ & & \\ [6pt]
\hline
$i=2$ & $t=1$ & $\sigma_{vrr}^{(P)} + \sigma_{wrr}^{(P)}$ & $\sigma_{vrr}^{(P)}$ & $\sigma_{rr}^{(P)}$ & \\
    & $t=2$ & $\sigma_{vrr}^{(P)}$ & $\sigma_{vrr}^{(P)} + \sigma_{wrr}^{(P)}$ & $\sigma_{vrr}^{(P)} + \sigma_{urr}^{(P)}$ & $\sigma_{rr}^{(P)}$ \\
\hline
\end{tabular}}
\end{center}
\end{footnotesize}
\end{table}

\subsection{Estimation}
\label{sec:methods.estimation}

Maximum likelihood estimation of model (\ref{eq:RPmodel}) can be carried out using numerical quadrature to integrate out the random effects.  Although there are functions available for fitting multivariate random effects probit models in standard software (e.g. \texttt{cmp} in Stata), these are typically not computationally feasible for high-dimensional multivariate outcomes combined with large sample sizes.  Model (\ref{eq:RCmodel}) cannot be estimated via maximum likelihood in existing software because the inclusion of the additional random effect $w_{rtj}^{(P)}$ leads to a non-hierarchical structure.  MCMC software is a natural alternative, but estimation is slow and imposing the unit constraints on the variances in the covariance matrices of the time-varying residuals ($\bm e_{ti}^{(C)}$ and $\bm e_{tij}^{(P)}$) is not straightforward.  We therefore propose an efficient Gibbs sampling algorithm for a class of multivariate random effect probit models, including the models of (\ref{eq:RPmodel}) and (\ref{eq:RCmodel}) as special cases.

%%
%% Edited by Siliang on 6 Sep 2021
%%
Omitting all superscripts and subscripts, the sampling procedure alternates between sampling of the latent variables (latent responses $\bm y^*$ and random effects) and of the model parameters (coefficients, random effect covariance matrices, and correlations of the time-varying residuals). There are two challenges for efficient estimation of the multivariate random effect probit models with residual variance constraints. The first is efficient sampling of $\bm y^*$ from the truncated multivariate normal distribution, which is handled by a Choleksy factorization similar to the GHK algorithm \citep{geweke.1991,hajivassiliou.etal.1996,keane.1993}. The second challenge is estimation of the correlation matrix $\bm \Sigma_e$ under the positive definite constraint. We use a random-walk Metropolis sampler and a special joint uniform prior for the correlation parameters to carry out the estimation within these constraints. Full details of the estimation, including derivations and proofs, are provided in the supplementary materials. The estimation algorithm is implemented in the open source R package \texttt{mvreprobit} \citep{mvreprobit}.
%The latent responses $\bm y^*$ are sampled from a truncated multivariate normal distribution, the random effects and coefficients are sampled from full multivariate normals, and the random effect covariance matrices are sampled from inverse Wishart distributions. To account for the identification constraints on the variances in $\bm \Sigma_e$, the proposed algorithm includes a random-walk Metropolis Hastings step where the elements of $\bm \Sigma_e$ are sampled from a joint uniform distribution and each proposal is checked to ensure that the obtained $\bm \Sigma_e$ is positive semi-definite, i.e. a proper correlation matrix. Full details are provided in the supplementary materials. An R package has been developed and will be available online as open source.

\section{Data analysis}
\label{sec:analysis}

\subsection{Preliminaries}
\label{sec:analysis.prelim}

The joint random effects model for the four respondent-parent exchanges described in Section \ref{sec:methods.RPmodel} was fitted to the UKHLS data, including the covariates given in Table \ref{tab:descriptives} together with dummy variables for wave.  In preliminary analysis, we considered whether the effects of the presence and age of children differed for unpartnered and partnered respondents, for example to investigate whether lone-parent families were more likely to receive parental support than two-parent families.  We found no evidence of any such interaction effect.  The posterior means and standard deviations of the coefficients, random effect variances and within-individual correlations for the final specification of model (\ref{eq:RPmodel}) are given in Table \ref{tab:modelRP.coeff}.  These are based on pooling two parallel chains of 7,500 MCMC iterations, each using different starting values, after discarding a burn-in sample of 2,500.   Convergence was assessed using a range of graphical diagnostics and the potential scale reduction factor (PSRF) \citep{gelman.etal.2004}. Final PSRF estimates were close to 1 for all parameters, and increasing the chain length led to little change in the running means of the posterior estimates.

The more complex model of Section \ref{sec:methods.RCmodel}, with individual and time-invariant and time-varying couple effects, was fitted to data on the four types of respondent-child exchanges.  However, the estimates for random effect correlations between receipt of financial support and the other three outcomes displayed poor mixing and the running means of the posterior estimates of these parameters showed some substantial differences between chains.  The source of the problem was the small response probability (2.5\%) for receipt of financial help (in spite of the large sample size).  We therefore fitted the joint model to the other three outcomes and a separate model for receipt of financial help (with the same random effects structure but without allowing for random effect and residual correlations between receipt of financial help and the other outcomes).  Table \ref{tab:modelRC.coeff} shows results from the final specification of model (\ref{eq:RCmodel}), based on pooling two parallel chains of 50,000 MCMC iterations for the joint model (with a burn-in of 5,000) and of 100,000 iterations for the receipt of financial help model (burn-in of 10,000).

It is difficult to assess the strength of covariate effects from the coefficient estimates in Tables \ref{tab:modelRP.coeff} and \ref{tab:modelRC.coeff} because for multilevel generalised linear models the addition of random effects increases the scale of the underlying latent responses (see \citet[][Chapter 17]{snijders.bosker.2012} for the case of two-level models for binary responses). We therefore calculated predicted marginal response probabilities to illustrate the magnitude of covariate effects on each type of parent-child exchange and to facilitate comparisons across outcomes (see Tables \ref{tab:modelRP.prob} and \ref{tab:modelRC.prob}).  To obtain the mean prediction $\hat{\mbox{Pr}} (y_r=1 | x_k=c)$ for outcome $r$ and covariate $x_k$, we first compute a predicted probability for each person-wave observation based on the posterior means of the model parameters, setting $x_k=c$ and holding the other covariates at their observed values; we then take the average of the predictions across observations.  The random effects are integrated out, as described by \cite{bland.cook.2019}.

The models presented here include travel time between children and parents as a covariate.  However, as noted by other authors \citep[e.g.][]{heylen.etal.2012}, geographical proximity is likely to be endogenous with respect to exchanges of support because individuals may move closer to parents or children to facilitate provision or receipt of practical help in particular.  We find that the inclusion of distance has little impact on the significance or magnitude of covariate effects on giving or receiving financial support, which is consistent with the weak association between proximity and financial exchanges. The results from models without travel time are given in supplementary materials (Tables S3-S6) and in the interpretation below we comment on how the inclusion of proximity affects the significance and magnitude of effects on exchanges of practical help presented in Tables \ref{tab:modelRP.coeff}--\ref{tab:modelRC.prob}.  The discussion of the impact of travel time on our results is also based on ordered logit analysis of child-parent travel times, with standard errors corrected for individual clustering (results not shown).

In the following sections we examine research questions 1-4 from Section \ref{sec:previous.research}. The discussion of questions 1 and 2 are based on estimates of the coefficients of the covariates, $\boldsymbol\beta_r^{(C)}$ and $\boldsymbol\beta_r^{(P)}$, in the models described in Sections \ref{sec:methods.RPmodel} and \ref{sec:methods.RCmodel}, while questions 3 and 4 are investigated using estimates of the random effect variances and covariances.

\begin{table}
\caption{\label{tab:modelRP.coeff} Results from multivariate random effects probit model for exchanges with non-coresident parents from an adult child perspective. The estimates are posterior means from MCMC samples (and posterior standard deviations in parentheses).}
\begin{footnotesize}
\centering
\fbox{%
\begin{tabular}{l d{2.3}@{}l r d{2.3}@{}l r d{2.3}@{}l r d{2.3}@{}l r}
\hline
&   \multicolumn{3}{c}{To parents:} & \multicolumn{3}{c}{To parents:} & \multicolumn{3}{c}{From parents:} & \multicolumn{3}{c}{From parents:} \\
&   \multicolumn{3}{c}{practical} & \multicolumn{3}{c}{financial} & \multicolumn{3}{c}{practical} & \multicolumn{3}{c}{financial} \\
&   \multicolumn{3}{c}{$(r=1)$} & \multicolumn{3}{c}{$(r=2)$} & \multicolumn{3}{c}{$(r=3)$} & \multicolumn{3}{c}{$(r=4)$} \\
\cmidrule(lr){2-4} \cmidrule(lr){5-7} \cmidrule(lr){8-10} \cmidrule(lr){11-13}
Variable & \multicolumn{2}{c}{Est.} & \multicolumn{1}{c}{(SD)} & \multicolumn{2}{c}{Est.} & \multicolumn{1}{c}{(SD)} & \multicolumn{2}{c}{Est.} & \multicolumn{1}{c}{(SD)} & \multicolumn{2}{c}{Est.} & \multicolumn{1}{c}{(SD)} \\
\hline
\multicolumn{13}{l}{\emph{Coefficients of explanatory variables, $\boldsymbol\beta_r^{(C)}$}} \\
%template
%& -0.000 & $^*$ & (0.000) & -0.000 & $^*$ & (0.000) & -0.000 & $^*$ & (0.000) & -0.000 & $^*$ & (0.000) \\
Age (years)$^a$
& -0.013 & $^*$ & (0.002) &  0.006 &      & (0.003) & -0.049 & $^*$ & (0.002) & -0.058 & $^*$ & (0.003) \\
Female
&  0.157 & $^*$ & (0.022) & -0.086 & $^*$ & (0.033) &  0.277 & $^*$ & (0.021) &  0.171 & $^*$ & (0.025) \\
Ethnicity (ref=White) & & & & & & & & \\
~~Asian/ Asian British
&  0.632 & $^*$ & (0.045) &  0.971 & $^*$ & (0.060) & -0.280 & $^*$ & (0.043) & -0.213 & $^*$ & (0.054) \\
~~Black/ Black British
&  0.293 & $^*$ & (0.064) &  1.185 & $^*$ & (0.080) &  0.038 &      & (0.063) &  0.084 &      & (0.070) \\
~~Other
&  0.170 & $^*$ & (0.074) &  0.569 & $^*$ & (0.099) & -0.020 &      & (0.068) & -0.120 &      & (0.080) \\
Coresident partner
& -0.070 & $^*$ & (0.024) &  0.020 &      & (0.039) & -0.478 & $^*$ & (0.024) & -0.420 & $^*$ & (0.027) \\
Long-term illness
& -0.151 & $^*$ & (0.026) &  0.002 &      & (0.040) &  0.016 &      & (0.027) &  0.137 & $^*$ & (0.030) \\
Post-school education
& -0.034 &      & (0.023) &  0.160 & $^*$ & (0.035) &  0.113 & $^*$ & (0.021) & -0.036 &      & (0.025) \\
Unemp./ econ. inactive
&  0.018 &      & (0.024) & -0.064 &      & (0.039) & -0.039 &      & (0.024) &  0.108 & $^*$ & (0.027) \\
Log annual hh inc.
& -0.034 & $^*$ & (0.012) &  0.100 & $^*$ & (0.020) &  0.012 &      & (0.012) & -0.142 & $^*$ & (0.012) \\
Home owner
&  0.021 &      & (0.024) & -0.088 & $^*$ & (0.039) &  0.143 & $^*$ & (0.024) & -0.319 & $^*$ & (0.026) \\
\multicolumn{13}{l}{Child coresidence status (ref=none)}   \\
~~Cores. only$^b$
& -0.188 & $^*$ & (0.035) & -0.046 &      & (0.059) &  0.707 & $^*$ & (0.032) &  0.052 &      & (0.037) \\
~~Cores. and non-cores$^b$
& -0.125 & $^*$ & (0.044) & -0.049 &      & (0.074) &  0.599 & $^*$ & (0.042) &  0.132 & $^*$ & (0.051) \\
~~Non-cores. only
&  0.044 &      & (0.037) & -0.129 & $^*$ & (0.059) & -0.027 &      & (0.038) &  0.064 &      & (0.047) \\
\multicolumn{13}{l}{Age of youngest coresident child (ref=$<2$ yrs)}  \\
~~2--4 yrs
&  0.069 & $^*$ & (0.032) & -0.002 &      & (0.056) &  0.036 &      & (0.030) &  0.049 &      & (0.035) \\
~~5--10 yrs
&  0.074 & $^*$ & (0.034) & -0.013 &      & (0.060) & -0.003 &      & (0.032) &  0.088 & $^*$ & (0.038) \\
~~11--16 yrs
&  0.121 & $^*$ & (0.041) & -0.056 &      & (0.067) & -0.459 & $^*$ & (0.038) &  0.042 &      & (0.047) \\
~~$>$ 16 yrs
&  0.177 & $^*$ & (0.046) & -0.112 &      & (0.073) & -0.796 & $^*$ & (0.045) &  0.025 &      & (0.053) \\
\multicolumn{13}{l}{Number of siblings (ref=none)} \\
~~1
& -0.060 &      & (0.034) & -0.160 & $^*$ & (0.052) &  0.125 & $^*$ & (0.033) & -0.084 & $^*$ & (0.038) \\
~~$\geq$ 2
& -0.079 & $^*$ & (0.032) & -0.058 &      & (0.051) & -0.124 & $^*$ & (0.032) & -0.268 & $^*$ & (0.037) \\
Age of oldest parent (yrs)$^a$
&  0.039 & $^*$ & (0.002) &  0.008 & $^*$ & (0.003) & -0.012 & $^*$ & (0.002) &  0.020 & $^*$ & (0.002) \\
Parent age squared $\times 10^{-1}$
&  0.008 & $^*$ & (0.001) &  0.011 & $^*$ & (0.001) & -0.012 & $^*$ & (0.001) & -0.002 &      & (0.001) \\
$\geq$ 1 parent lives alone
&  0.627 & $^*$ & (0.020) &  0.504 & $^*$ & (0.032) & -0.169 & $^*$ & (0.020) &  0.014 &      & (0.024) \\
\multicolumn{13}{l}{Travel time to closest parent (ref=$<$ 15 mins)}   \\
~~15--30 mins
& -0.430 & $^*$ & (0.023) & -0.132 & $^*$ & (0.038) & -0.317 & $^*$ & (0.022) & -0.109 & $^*$ & (0.027) \\
~~30--60 mins
& -0.851 & $^*$ & (0.030) & -0.248 & $^*$ & (0.048) & -0.663 & $^*$ & (0.029) & -0.134 & $^*$ & (0.035) \\
~~1--2 hrs
& -1.335 & $^*$ & (0.034) & -0.306 & $^*$ & (0.056) & -1.089 & $^*$ & (0.033) & -0.168 & $^*$ & (0.039) \\
~~$>$ 2 hrs
& -1.989 & $^*$ & (0.034) & -0.242 & $^*$ & (0.046) & -1.763 & $^*$ & (0.032) & -0.254 & $^*$ & (0.035) \\
Constant
&  0.475 &      & (0.121) & -3.700 &      & (0.211) & -0.107 &      & (0.123) &  0.384 &      & (0.129) \\
\emph{Random effect variances} $\sigma_{urr}^{(C)}$
&  1.465 &      & (0.121) &  1.849 &      & (0.081) &  1.105 &      & (0.036) &  1.332 &      & (0.051) \\
\emph{Within-individual correlations}
&  0.594 &      & (0.007) &  0.649 &      & (0.010) &  0.525 &      & (0.008) &  0.569 &      & (0.010) \\
\hline
\multicolumn{13}{p{\textwidth}}{$^*$95\% credible interval does not include zero; $^a$Respondent age is centred around 40 and parental age around 70; squared respondent age is also included in the model, but its effects are negligible and non-significant. $^b$Contrasts 1+ coresident child where youngest is age $<$2 years versus no children. $^c$Effects of age of youngest child among respondents with coresident children.}
\end{tabular}}
\end{footnotesize}
\end{table}

%%%%%%%%%%%%%%%%%%%%%%%%%%%%%%%%%%%%%%%

\begin{table}
\caption{\label{tab:modelRC.coeff} Results from multivariate random effects probit model for exchanges with non-coresident children from a parental perspective. The estimates are posterior means from MCMC samples (and posterior standard deviations in parentheses).}
\begin{footnotesize}
\centering
\fbox{%
\begin{tabular}{l d{2.3}@{}l r d{2.3}@{}l r d{2.3}@{}l r d{2.3}@{}l r}
\hline
&   \multicolumn{3}{c}{To children:} & \multicolumn{3}{c}{To children:} & \multicolumn{3}{c}{From children:} & \multicolumn{3}{c}{From children:} \\
&   \multicolumn{3}{c}{practical} & \multicolumn{3}{c}{financial} & \multicolumn{3}{c}{practical} & \multicolumn{3}{c}{financial} \\
&   \multicolumn{3}{c}{$(r=1)$} & \multicolumn{3}{c}{$(r=2)$} & \multicolumn{3}{c}{$(r=3)$} & \multicolumn{3}{c}{$(r=4)$} \\
\cmidrule(lr){2-4} \cmidrule(lr){5-7} \cmidrule(lr){8-10} \cmidrule(lr){11-13}
Variable & \multicolumn{2}{c}{Est.} & \multicolumn{1}{c}{(SD)} & \multicolumn{2}{c}{Est.} & \multicolumn{1}{c}{(SD)} & \multicolumn{2}{c}{Est.} & \multicolumn{1}{c}{(SD)} & \multicolumn{2}{c}{Est.} & \multicolumn{1}{c}{(SD)} \\
\hline
\multicolumn{13}{l}{\emph{Coefficients of explanatory variables, $\boldsymbol\beta_r^{(P)}$}} \\
%template
%& -0.000 & $^*$ & (0.000) & -0.000 & $^*$ & (0.000) & -0.000 & $^*$ & (0.000) & -0.000 & $^*$ & (0.000) \\
Age (years)
&  0.058 & $^*$ & (0.005) & -0.036 & $^*$ & (0.005) &  0.018 & $^*$ & (0.005) & -0.009 & $^*$ & (0.010) \\
Age squared $\times 10^{-1}$
& -0.030 & $^*$ & (0.001) & -0.003 & $^*$ & (0.001) &  0.004 & $^*$ & (0.001) &  0.000 &      & (0.002) \\
Female
&  0.372 & $^*$ & (0.026) & -0.298 & $^*$ & (0.025) &  0.642 & $^*$ & (0.029) &  0.368 & $^*$ & (0.067) \\
\multicolumn{13}{l}{Ethnicity (ref=White)} \\
~~Asian/ Asian British
& -0.688 & $^*$ & (0.084) & -0.957 & $^*$ & (0.093) &  0.389 & $^*$ & (0.082) &  1.638 & $^*$ & (0.176) \\
~~Black/ Black British
& -0.087 &      & (0.094) & -0.105 &      & (0.098) &  0.239 & $^*$ & (0.092) &  1.639 & $^*$ & (0.184) \\
~~Other
&  0.116 &      & (0.126) &  0.037 &      & (0.126) &  0.272 & $^*$ & (0.127) &  0.826 & $^*$ & (0.227) \\
Coresident partner
&  0.198 & $^*$ & (0.035) & -0.006 &      & (0.036) & -0.756 & $^*$ & (0.038) & -0.825 & $^*$ & (0.093) \\
Long-term illness
& -0.311 & $^*$ & (0.028) &  0.011 &      & (0.028) &  0.364 & $^*$ & (0.028) &  0.397 & $^*$ & (0.069) \\
Post-school education
&  0.186 & $^*$ & (0.033) &  0.532 & $^*$ & (0.034) & -0.162 & $^*$ & (0.034) & -0.110 &      & (0.078) \\
Unemp./ econ. inactive
&  0.175 & $^*$ & (0.034) & -0.256 & $^*$ & (0.034) &  0.122 & $^*$ & (0.034) &  0.198 & $^*$ & (0.082) \\
Log annual hh inc.
&  0.003 &      & (0.019) &  0.194 & $^*$ & (0.021) & -0.020 &      & (0.019) & -0.145 & $^*$ & (0.036) \\
Home owner
&  0.378 & $^*$ & (0.040) &  0.429 & $^*$ & (0.041) & -0.232 & $^*$ & (0.039) & -0.712 & $^*$ & (0.090) \\
Any coresident children
& -0.083 & $^*$ & (0.038) & -0.208 & $^*$ & (0.038) & -0.004 &      & (0.038) &  0.253 & $^*$ & (0.082) \\
\multicolumn{13}{l}{No. non-cores. children (ref=1)} \\
~~2
&  0.310 & $^*$ & (0.037) &  0.102 & $^*$ & (0.037) &  0.173 & $^*$ & (0.038) &  0.146 &      & (0.085) \\
~~$\geq$ 3
&  0.400 & $^*$ & (0.042) &  0.174 & $^*$ & (0.041) &  0.425 & $^*$ & (0.042) &  0.396 & $^*$ & (0.092) \\
Has a surviving parent
& -0.163 & $^*$ & (0.035) & -0.064 &      & (0.034) & -0.314 & $^*$ & (0.036) & -0.439 & $^*$ & (0.090) \\
\multicolumn{13}{l}{Travel time to child in most contact (ref=$<$ 15 mins)}  \\
~~15--30 mins
& -0.515 & $^*$ & (0.034) & -0.117 & $^*$ & (0.033) & -0.425 & $^*$ & (0.032) & -0.092 &      & (0.074) \\
~~30--60 mins
& -0.999 & $^*$ & (0.047) & -0.159 & $^*$ & (0.043) & -0.834 & $^*$ & (0.046) & -0.071 &      & (0.092) \\
~~1--2 hrs
& -1.540 & $^*$ & (0.058) & -0.029 &      & (0.047) & -1.318 & $^*$ & (0.056) & -0.333 & $^*$ & (0.115) \\
~~$>$ 2 hrs
& -2.401 & $^*$ & (0.071) & -0.104 & $^*$ & (0.043) & -2.225 & $^*$ & (0.070) & -0.409 & $^*$ & (0.110) \\
Constant
&  0.306 &      & (0.196) & -2.007 &      & (0.209) & -0.465 &      & (0.195) & -2.522 &      & (0.417) \\
\multicolumn{13}{l}{\emph{Random effect variances}} \\
Time-invariant couple $\sigma_{vrr}^{(P)}$
&  1.364 &      & (0.075) &  1.444 &      & (0.073) &  1.233 &      & (0.070) &  1.659 &      & (0.255) \\
Time-varying couple $\sigma_{wrr}^{(P)}$
&  1.373 &      & (0.107) &  1.257 &      & (0.089) &  1.301 &      & (0.102) &  1.991 &      & (0.461) \\
Time-invariant ind. $\sigma_{urr}^{(P)}$
&  0.786 &      & (0.073) &  0.706 &      & (0.061) &  0.673 &      & (0.064) &  0.732 &      & (0.287) \\
\multicolumn{13}{l}{\emph{Intra-cluster correlations}} \\
Within-individual
&  0.475 &      & (0.008) &  0.488 &      & (0.008) &  0.453 &      & (0.008) &  0.444 &      & (0.022) \\
Within-couple
&  0.605 &      & (0.010) &  0.612 &      & (0.010) &  0.602 &      & (0.011) &  0.680 &      & (0.029) \\
\hline
\multicolumn{13}{p{\textwidth}}{$^*$95\% credible interval does not include zero.}
\end{tabular}}
\end{footnotesize}
\end{table}

%%%%%%%%%%%%%%%%%%%%%%%%%%%%%%%%%%%%%%%

\begin{table}
\caption{\label{tab:modelRP.prob} Predicted marginal probabilities of giving and receiving practical and financial help to/from parents (from an adult child perspective), calculated from model of Table \ref{tab:modelRP.coeff}.  }
\begin{footnotesize}
\centering
\fbox{%
\begin{tabular}{l cccc}
\hline
Respondent (child) & To parents: & To parents: & From parents: & From parents: \\
characteristics & practical & financial & practical & financial \\
& $(r=1)$ & $(r=2)$ & $(r=3)$ & $(r=4)$ \\
\hline
%template
%&  .000   &   .000    &   .000    &   .000 \\
\multicolumn{4}{l}{Age} \\
~~30 years
&  .465   &   .056    &   .472    &   .241 \\
%~~40
%&  .433   &   .062    &   .357    &   .143 \\
~~50
&  .410   &   .065    &   .257    &   .076 \\
\multicolumn{4}{l}{Gender} \\
~~Male
&  .412   &   .065    &   .322    &   .122 \\
~~Female
&  .446   &   .059    &   .377    &   .144 \\
\multicolumn{4}{l}{Ethnicity} \\
~~Asian or Asian British
&  .560   &   .146    &   .303    &   .110 \\
~~Black or Black British
&  .486   &   .176    &   .365    &   .148 \\
~~White
&  .422   &   .053    &   .358    &   .137 \\
~~Other
&  .459   &   .099    &   .354    &   .133 \\
\multicolumn{4}{l}{Partnership status} \\
~~Unpartnered
&  .443   &   .061    &   .429    &   .179 \\
~~Partner
&  .428   &   .062    &   .332    &   .120 \\
\multicolumn{4}{l}{Has long-term illness} \\
~~No
&  .436   &   .062    &   .354    &   .133 \\
~~Yes
&  .404   &   .062    &   .357    &   .151 \\
\multicolumn{4}{l}{Has post-school education} \\
~~No
&  .435   &   .057    &   .344    &   .137 \\
~~Yes
&  .428   &   .068    &   .367    &   .132 \\
\multicolumn{4}{l}{Employment status} \\
~~Unemployed/econ. inactive
&  .435   &   .058    &   .348    &   .146 \\
~~Employed
&  .431   &   .063    &   .356    &   .131 \\
\multicolumn{4}{l}{Equivalised household income} \\
~~10th percentile
&  .436   &   .057    &   .353    &   .145 \\
%~~25th
%&  .434   &   .059    &   .354    &   .139 \\
~~50th
&  .431   &   .062    &   .355    &   .132 \\
%~~75th
%&  .429   &   .064    &   .355    &   .127 \\
~~90th
&  .396   &   .103    &   .367    &   .062 \\
\multicolumn{4}{l}{Housing tenure} \\
~~Social/private rent or other
&  .429   &   .066    &   .335    &   .162 \\
~~Own home
&  .433   &   .060    &   .364    &   .119 \\
\multicolumn{4}{l}{Presence/age of children} \\
~~None
&  .441   &   .067    &   .292    &   .126 \\
~~Coresident, youngest $<$ 2 yrs
&  .401   &   .064    &   .443    &   .132 \\
~~Coresident, youngest 2--4
&  .415   &   .063    &   .451    &   .139 \\
~~Coresident, youngest 5--10
&  .417   &   .063    &   .442    &   .144 \\
~~Coresident, youngest 11--16
&  .427   &   .060    &   .343    &   .138 \\
~~Coresident, youngest $>$ 16
&  .439   &   .056    &   .275    &   .135 \\
~~Cores. and non-cores, youngest $>$ 16
&  .453   &   .056    &   .255    &   .146 \\
~~Non-coresident only
&  .451   &   .058    &   .287    &   .134 \\
\multicolumn{4}{l}{Number of siblings} \\
~~0
&  .446   &   .067    &   .360    &   .160 \\
~~1
&  .433   &   .056    &   .385    &   .148 \\
~~$\geq$ 2
&  .429   &   .063    &   .335    &   .124 \\
\multicolumn{4}{l}{Age of oldest parent}  \\
%~~60 years
%&  .337   &   .052    &   .380    &   .119 \\
~~70
&  .403   &   .050    &   .380    &   .146 \\
~~80
&  .507   &   .063    &   .329    &   .172 \\
~~90
&  .643   &   .098    &   .236    &   .196 \\
\multicolumn{4}{l}{At least one parent lives alone} \\
~~No
&  .381   &   .048    &   .366    &   .134 \\
~~Yes
&  .520   &   .084    &   .332    &   .136 \\
\multicolumn{4}{l}{Travel time to nearest parent} \\
~~$<$15 mins
&  .578   &   .071    &   .464    &   .147 \\
~~15--30 mins
&  .476   &   .061    &   .392    &   .133 \\
~~30--60 mins
&  .377   &   .054    &   .318    &   .130 \\
~~1--2 hours
&  .274   &   .050    &   .235    &   .125 \\
~~$>$ 2 hours
&  .161   &   .054    &   .131    &   .115 \\
%\hline
%Overall
%&  .432   &   .062    &   .354    &   .135 \\
\hline
\end{tabular}}
\end{footnotesize}
\end{table}

%%%%%%%%%%%%%%%%%%%%%%%%%%%%%%%%%%%%%%%

\begin{table}
\caption{\label{tab:modelRC.prob} Predicted marginal probabilities of giving and receiving practical and financial help to/from adult children (from a parental perspective), calculated from model of Table \ref{tab:modelRC.coeff}.  }
\begin{footnotesize}
\centering
\fbox{%
\begin{tabular}{l cccc}
\hline
Respondent (parent) & To children: & To children: & From children: & From children: \\
characteristics & practical & financial & practical & financial \\
& $(r=1)$ & $(r=2)$ & $(r=3)$ & $(r=4)$ \\
\hline
%template
%&  .000   &   .000    &   .000    &   .000 \\
\multicolumn{4}{l}{Age} \\
~~50 years
&  .633   &   .401    &   .285    &   .029 \\
%~~60
%&  .609   &   .323    &   .328    &   .027 \\
~~70
&  .500   &   .244    &   .386    &   .024 \\
\multicolumn{4}{l}{Gender} \\
~~Male
&  .484   &   .316    &   .305    &   .020 \\
~~Female
&  .545   &   .270    &   .405    &   .029 \\
\multicolumn{4}{l}{Ethnicity} \\
~~Asian or Asian British
&  .410   &   .166    &   .422    &   .085 \\
~~Black or Black British
&  .509   &   .278    &   .398    &   .085 \\
~~White
&  .523   &   .294    &   .360    &   .021 \\
~~Other
&  .542   &   .300    &   .403    &   .044 \\
\multicolumn{4}{l}{Partnership status} \\
~~Unpartnered
&  .496   &   .290    &   .449    &   .038 \\
~~Partner
&  .528   &   .289    &   .325    &   .018 \\
\multicolumn{4}{l}{Has long-term illness} \\
~~No
&  .535   &   .289    &   .344    &   .022 \\
~~Yes
&  .484   &   .290    &   .402    &   .031 \\
\multicolumn{4}{l}{Has post-school education} \\
~~No
&  .510   &   .264    &   .370    &   .026 \\
~~Yes
&  .541   &   .348    &   .345    &   .024 \\
\multicolumn{4}{l}{Employment status} \\
~~Unemployed/econ. inactive
&  .530   &   .273    &   .369    &   .027 \\
~~Employed
&  .501   &   .312    &   .350    &   .022 \\
\multicolumn{4}{l}{Equivalised household income} \\
~~10th percentile
&  .519   &   .271    &   .365    &   .027 \\
%~~25th
%&  .519   &   .280    &   . 364   &   .026 \\
~~50th
&  .519   &   .290    &   .363    &   .025 \\
%~~75th
%&  .519   &   .300    &   .362    &   .024 \\
~~90th
&  .521   &   .449    &   .348    &   .012 \\
\multicolumn{4}{l}{Housing tenure} \\
~~Social/private rent or other
&  .471   &   .240    &   .391    &   .038 \\
~~Own home
&  .533   &   .304    &   .355    &   .020 \\
\multicolumn{4}{l}{Has coresident children} \\
~~No
&  .522   &   .298    &   .363    &   .024 \\
~~Yes
&  .509   &   .266    &   .363    &   .030 \\
{Number of non-coresident children} \\
~~1
&  .476   &   .274    &   .329    &   .021 \\
~~2
&  .527   &   .290    &   .356    &   .024 \\
~~$\geq$ 3
&  .541   &   .301    &   .396    &   .030 \\
\multicolumn{4}{l}{Has surviving parent} \\
~~No
&  .527   &   .293    &   .377    &   .029 \\
~~Yes
&  .501   &   .283    &   .328    &   .019 \\
\multicolumn{4}{l}{Travel time to child in most contact} \\
~~$<$15 mins
&  .639   &   .299    &   .458    &   .028 \\
~~15--30 mins
&  .553   &   .281    &   .385    &   .026 \\
~~30--60 mins
&  .468   &   .274    &   .318    &   .026 \\
~~1--2 hours
&  .375   &   .294    &   .246    &   .021 \\
~~$>$ 2 hours
&  .242   &   .283    &   .138    &   .019 \\
%\hline
%Overall
%&  .519   &   .289    &   .363    &   .025 \\
\hline
\end{tabular}}
\end{footnotesize}
\end{table}

\subsection{Predictors of parent-child exchanges of practical and financial support}
\label{sec:analysis.predictors}

As noted in Section \ref{sec:previous.research}, one way to understand the results is to think about the covariates as characterising the capacity of one actor in the parent-child relationship to provide help, the needs of the other partner to receive help, and the operation of social or cultural norms.  We use this framework in the following discussion, starting with parents as the potential recipients and then children.

\subsubsection*{Child-to-parent support (research question 1)}

We begin by examining how provision of help by adult children to their parents varies by individual characteristics.  We study child-to-parent exchanges from the perspective of both child respondents ($r=1$ and $r=2$ in Tables \ref{tab:modelRP.coeff} and \ref{tab:modelRP.prob}) and parent respondents ($r=3$ and $r=4$ in Tables \ref{tab:modelRC.coeff} and \ref{tab:modelRC.prob}) in order to consider the effects of the characteristics of each generation. However, when making comparisons across generations it is important to note that the two perspectives are asymmetric: a child respondent reports collectively on exchanges with both parents, and a parent with more than one child reports on all children collectively. The direction and statistical significance of covariate effects can be obtained from Tables \ref{tab:modelRP.coeff} and \ref{tab:modelRC.coeff}, and the predicted probabilities in Tables \ref{tab:modelRP.prob} and \ref{tab:modelRC.prob} show their magnitude. The following discussion of the effects of child characteristics is based on Tables \ref{tab:modelRP.coeff} and \ref{tab:modelRP.prob}; the discussion of the effects of parent characteristics is based largely on Tables \ref{tab:modelRC.coeff} and \ref{tab:modelRC.prob}, with the exception of age of oldest parent and whether any parent lives alone which were asked of child respondents. 

In terms of \textbf{needs}, we find that the probability that a parent receives practical and financial help increases with parental age (after controlling for child age), and there is a moderate effect of parental health such that parents with a limiting illness are more likely to be recipients of both practical and financial help. Parents without a partner are substantially more likely to receive both practical and financial support from their children and, consistent with this, we also find that children are more likely to help parents when at least one parent is living alone. However, the presence of an older generation, that is, the parent him or herself having a surviving parent, is negatively associated with the parent's probability of receiving either practical or financial help from their children.

Socio-economic markers of potential need among parents suggest that those with greater needs are more likely to receive help from children. Non-employed parents are more likely than those in employment to receive practical and financial help, as are private or social renters compared to owner-occupiers. Parents with lower levels of education are more likely to receive practical help than parents with higher levels of education, and lower income parents are more likely to receive financial help (although the effects are small).  The effect of education strengthens when proximity is omitted (\cref{tab:modelRC.nodist.prob}): the negative association between receipt of practical help and higher education is partly explained by a tendency of more educated parents to live farther from their children.

In terms of \textbf{capacities}, as expected, parents who live closer to their children have a substantially higher probability of receiving practical help; the effects of distance on financial help are in the same direction but much weaker. Variations in receipt of help from a child are also associated with the child's health and family and household composition. Limiting long-term illness of the child is negatively associated with provision of practical help to their parents, although the effect is weak. Those with young coresident children are less likely than those without children to give practical help to their parents, but the effects are small and diminish with increasing age of the youngest child, and there is no effect on financial help.  The higher probability of giving practical help for respondents with no coresident children emerges after accounting for proximity (\cref{tab:modelRP.nodist.prob}) due to a tendency for these respondents to live further from their parents than those with children. Perhaps surprisingly, however, children with a coresident partner are less likely than unpartnered children to give practical help to parents.

There is some evidence that adult children with at least two siblings are less likely to give practical help to their parents than those with no or one sibling, possibly because the other children may provide support.  This effect is consistent with the finding that the more non-coresident children a parent has, the greater their chance of receiving support, and is also in line with previous studies.

Turning to socio-economic factors, we might expect children in better economic circumstances to have greater capacity to provide help to parents. In fact, there is a somewhat mixed picture. There is little evidence of an effect of a child's employment status on giving help to parents.  Higher income children are more likely to give financial help to their parents, but the effects are small. Higher child education is also positively associated with parental receipt of financial help. On the other hand, children who own their own home (the majority with a mortgage) are slightly less likely than private or social renters to give financial help.  The effects of all socio-economic characteristics on provision of practical support to parents are stronger before controlling for proximity (\cref{tab:modelRP.nodist.prob}) because non-employment, post-school education, higher income and renting are all associated with longer travel times between children and parents.

Associations between giving or receiving help and gender and ethnicity can be interpreted as indicators of \textbf{social norms}. Women are less likely than men to give financial help to parents, but more likely to give practical help, which may suggest a tendency for women to substitute practical help for financial help, and vice versa for men (research question 4).  Mothers are more likely than fathers to receive either form of help. From both child and parent viewpoints, we find that White children are less likely than those from other ethnic groups to give financial or practical help to their parents, with the largest variations by ethnicity observed for financial help.  If controls for proximity are excluded, the probability of giving practical help to parents decreases among Black/Black British and Other children (\cref{tab:modelRP.nodist.prob,tab:modelRC.nodist.prob}), which is explained by their tendency to live farther from their parents than the other ethnic groups, thus inhibiting exchanges of practical help. However, Black/Black British children remain more likely than White children (and less likely than Asian/Asian British children) to provide practical help to parents regardless of whether proximity is controlled.

\subsubsection*{Parent-to-child support (research question 2)}

We next consider help given by parents to their non-coresident adult children, again from the perspective of the child ($r=3$ and $r=4$ in Tables \ref{tab:modelRP.coeff} and \ref{tab:modelRP.prob}) and the parent ($r=1$ and $r=2$ in Tables \ref{tab:modelRC.coeff} and \ref{tab:modelRC.prob}).

In terms of the \textbf{needs} of the children, the probability of giving practical or financial help to children decreases with child age, possibly reflecting increasing independence. Children with a long-term limiting illness are more likely to receive financial help from their parents, while the positive association with receipt of practical help does not reach statistical significance.  There is an increased probability that a child receives practical and financial help from parents if the child is unpartnered. Respondents with young coresident children (and no non-coresident children) are more likely than those without children to receive practical help from parents, but the probability of receiving practical help drops markedly when the youngest child reaches secondary school age; there is no effect of presence or age of coresident children on receipt of financial assistance.  The effect of having children on receipt of practical support from parents is stronger before accounting for proximity (\cref{tab:modelRP.nodist.prob}) due to parents and children living closer together when there are grandchildren.

Children not in work are more likely to receive financial help, and there are weak income effects in the expected direction: children have an increased probability of receiving financial support if their own income is low. Effects of housing tenure on financial support are in the same direction but stronger, with a higher chance of receiving such assistance among children who are renters rather than owners, although children who rent their home are less likely than owners to receive practical support. Surprisingly, higher education is associated with a higher chance of receiving practical help, after controlling for other characteristics.

In terms of the \textbf{capacity} of parents to provide support, geographical proximity once again plays an important role in relation to practical support given to children, and has a weak effect on provision of financial assistance. Parents with a limiting long-standing health condition are less likely to provide practical support but there is an increased probability that a child receives practical help from parents if the parent has a partner and neither parent is living alone. Children with one other sibling are less likely than only children or those with two or more children to receive practical support from parents, while the probability of receiving financial support declines with the number of siblings.  From a parental perspective, the probability of giving practical or financial help to non-coresident children increases with family size, but a parent has a lower probability of giving practical help to their children if they have a surviving parent.

Children are more likely to receive financial help if their parent is employed, if their parents’ income is high, and if their parents have higher levels of education. Effects of housing tenure on financial support are in the same direction but stronger, with a higher chance of receiving such assistance among{} children whose parents are home-owners.  Owner-occupier parents are also more likely than renters to give practical support to their children.  As for support given to parents, the effects of socio-economic characteristics on provision of practical support to children are in general stronger before accounting for proximity (\cref{tab:modelRC.nodist.prob}).

Turning to \textbf{social norms}, after adjusting for child age, older parental age is negatively associated with provision of practical help to children and positively associated with provision of financial help.  Female children are more likely than men to receive either form of support, and mothers are more likely than fathers to give practical support but less likely to give financial assistance.  Ethnic differences in giving support to children are generally smaller than for exchanges in the other direction. Asian and Asian British children are the least likely of all ethnic groups to receive financial or practical help from their parents; taken with their higher propensity to give help to parents, this suggests exchanges of help tend to be upwards for this group. This finding remains when proximity is excluded (\cref{tab:modelRP.nodist.prob,tab:modelRC.nodist.prob}).

\subsection{Correlations between exchanges}
\label{sec:correlations}

Table \ref{tab:models.corr} shows the correlations between exchanges before and after adjusting for covariates.   The unadjusted correlations are the tetrachoric correlations and the adjusted correlations are calculated from the MCMC chains for the random effect variances and covariances and residual correlations from models (\ref{eq:RPmodel}) and (\ref{eq:RCmodel}). For respondent-parent exchanges, the correlation between outcomes $r$ and $r^\prime$ is given by
\begin{equation}
\mbox{cor}(y_{rti}^{(C)*}, y_{r^\prime ti}^{(C)*} | \bm x_{ti}) = \frac{\sigma_{ur r^\prime}^{(C)}+\sigma_{er r^\prime}^{(C)}}
{\sqrt{(\sigma_{urr}^{(C)}+1)(\sigma_{ur^\prime r^\prime}^{(C)}+1)}}
\end{equation}
and for respondent-child(ren) exchanges, the correlation is
\begin{equation}
\mbox{cor}(y_{rtij}^{(P)*}, y_{r^\prime tij}^{(P)*} | \bm x_{tij}) = \frac{\sigma_{ur r^\prime}^{(P)}+\sigma_{vr r^\prime}^{(P)}+\sigma_{wr r^\prime}^{(P)}+\sigma_{er r^\prime}^{(P)}}
{\sqrt{(\sigma_{urr}^{(P)}+\sigma_{vrr}^{(P)}+\sigma_{wrr}^{(P)}+1)(\sigma_{ur^\prime r^\prime}^{(P)}+\sigma_{vr^\prime r^\prime}^{(P)}+\sigma_{wr^\prime r^\prime}^{(P)}+1)}}
\end{equation}
We can distinguish two types of correlation: (i) between giving and receiving the same or different forms of help, and (ii) between giving financial and giving practical help or between receiving financial and receiving practical help. We use correlations of type (i) to investigate the nature and extent of reciprocity or symmetry in child-parent exchanges (research question 3).  For example, a positive type (i) correlation implies that exchanges are symmetric, i.e. a tendency for the two generations to provide mutual support or for neither to support the other.  Correlations of type (ii) shed light on whether practical and financial support are substitutes or complements (research question 4). A positive type (ii) correlation implies, for example, that those whose unmeasured characteristics place them at an above-average propensity to give one form of support (either financial or practical) tend also to have an above-average propensity to give the other, while those who do not give one form tend also not to give the other.   Where one form of support is substituted for the other, for example if children tend to provide financial help in place of practical help, the corresponding type (ii) correlation would be negative.

\begin{table}
\caption{\label{tab:models.corr} Estimates of cross-outcome correlations from multivariate random effects probit models for exchanges between non-coresident adult children and parents from a child and parental perspective.  Unadjusted estimates are tetrachoric correlations (and SEs). Adjusted estimates and posterior means (and SDs) are computed from MCMC chains for models of Tables \ref{tab:modelRP.coeff} and \ref{tab:modelRC.coeff}.}
\begin{footnotesize}
\centering
\fbox{%
\begin{tabular}{l d{2.3} d{2.3} d{2.3} d{2.3}}
\hline
Type of correlation$^a$ & \multicolumn{2}{c}{Child reports} & \multicolumn{2}{c}{Parent reports} \\
\cmidrule(lr){2-3} \cmidrule(lr){4-5}
(all to/from parents$^b$) & \multicolumn{1}{c}{Unadjusted} & \multicolumn{1}{c}{Adjusted}
& \multicolumn{1}{c}{Unadjusted} & \multicolumn{1}{c}{Adjusted} \\
\hline
(i) To practical \& from practical
&   .322    &   .430    &   .347    &   .416 \\
&   (.006)  &   (.010)  &   (.006)  &   (.009) \\ [6pt]
(i) To practical \& from financial
&   .208    &   .274    &   .099    &   .260 \\
&   (.007)  &   (.016)  &   (.007)  &   (.011) \\ [6pt]
(i) To financial \& from practical
&   .043    &   .135    &   .081    &   \multicolumn{1}{c}{--$^c$} \\
&   (.010)  &   (.024)  &   (.014)  &    \\ [6pt]
(i) To financial \& from financial
&   .014    &   .057    &   -.039    &   \multicolumn{1}{c}{--$^c$} \\
&   (.012)  &   (.030)  &   (.015)  &    \\ [6pt]
\hline
(ii) To practical \& to financial
&   .482    &   .482    &   .422    &   \multicolumn{1}{c}{--$^c$} \\
&   (.008)  &   (.010)  &   (.012)  &    \\ [6pt]
(ii) From practical \& from financial
&   .469    &   .422    &   .432    &   .439 \\
&   (.006)  &   (.019)  &   (.006)  &   (.007) \\
\hline
\multicolumn{5}{p{0.9\textwidth}}{$^a$Type (i) correlations measure reciprocity of exchanges, and type (ii) correlations measure substitution between financial and practical help. $^b$Help given to and received from parents (from a child perspective) corresponds to help received from and given to children (from a parent’s perspective).  Note, however, that the correlations from the two sources may differ because children report on exchanges with $\leq$ 2 parent(s) while parents report on exchanges with $\geq$ 1 child(ren). $^c$Model for receipt of financial help from children estimated separately from model for exchanges of other type of support between respondents and children, so adjusted correlations denoted by -- are not estimated.}
\end{tabular} }
\end{footnotesize}
\end{table}

\subsubsection*{Reciprocity of child-parent exchanges (research question 3)}

Starting with correlations of type (i), we find that the residual correlations, adjusted for demographic and socio-economic characteristics (including age, household income and distance), are larger than the unadjusted correlations. This is due to the effects of age (of either the child or the parent) being in opposite directions for giving and receiving help, which dilutes these correlations when age is uncontrolled.   We focus on interpretation of the adjusted correlations (where these are estimated).  There is moderate reciprocity in practical help: 0.430 and 0.416 from a child and parental perspective respectively.  In contrast, there is little association between giving and receiving financial help.  A negative correlation would be expected if a respondent reports on financial transfers with only one parent or one child because if one member of the dyad is the `giver' this implies that the other is the `receiver'.  In UKHLS, however, respondents report on exchanges with both parents or all children; thus it is possible, for example, that a parent respondent receives financial support from one child and gives financial support to another.  The correlations between giving one form of support and receiving the other are all low.

The breakdown of the overall correlations into correlations between time-varying residuals and between individual random effects for pairs of outcomes reported by adult child respondents are shown in \cref{tab:corr.RP}.  For provision and receipt of practical help, the correlation is higher between time-varying residuals (0.49) than between time-invariant random effects (0.38), suggesting that symmetry in exchanges of practical help is slightly dominated by unmeasured factors specific to a given wave; these unmeasured factors may be transient characteristics of the child (respondent), their parents or the dyad. The picture is more complex for parent reports where the correlation can be further broken down into time-varying and time-invariant couple random effects (\cref{tab:corr.RC}), but there is a strong indication that symmetry in exchanges from a parental perspective is driven by unmeasured individual (time-varying or time-invariant) rather than couple characteristics; again, these individual factors may include characteristics of the parent, their children or the parent-child relationship.

\subsubsection*{Substitution or complementarity of practical and financial help (research question 4)}

The type (ii) correlations are all moderate and positive, which implies little evidence of substitution of one form of support for the other after accounting for respondent characteristics.  For both child and parent respondents, those whose propensity to give financial help is above the expected level (given their covariate values) tend also to have an above-average propensity to give practical help. The correlations between the time-varying and time-invariant residual components of child reports of support given are similar (\cref{tab:corr.RP}).  For parent reports the individual correlations (time-varying and time-invariant) are higher than the couple correlations (\cref{tab:corr.RC}), suggesting that complementarity in provision of practical and financial support from parents to children is dominated by transient or fixed characteristics of the respondent (or their children or parent-child relationship) rather than the shared characteristics of the respondent and their partner. For child respondents, a very similar pattern is found for support received: moderate and positive correlations between financial and practical help for both time-varying and individual residual components.  (Recall that correlations with receipt of financial support could not be estimated for parent respondents.)

Although from the cross-outcome correlations we conclude that, after controlling for covariates, financial and practical support are complementary rather than substitutes, a comparison of the signs of coefficients for giving practical and financial help (and for receiving practical and financial help) suggest that substitution effects operate within some subgroups. For example, higher income for child respondents is associated with a lower probability of giving help to parents but a higher probability of giving financial help, while child home-ownership is positively associated with receiving practical help and negatively associated with receiving financial help from parents (see Table \ref{tab:modelRP.coeff}). Based on parental reports of support given to children, we find that women and employed parents are more likely than their male and non-employed counterparts to give practical help but less likely to give financial help (Table \ref{tab:modelRC.coeff}).

\section{Discussion}
\label{sec:discussion}

In this paper, we study exchanges of financial and practical support between adult children and their parents using large-scale longitudinal data over an 8-year period. We employ multivariate random effects models to quantify reciprocity or mutuality of exchanges and the extent to which practical and financial assistance substitute for or complement each other. Exchanges are studied from the viewpoint of child and parent respondents to investigate the effects of the characteristics and circumstances of both generations.  When coresident partners report on exchanges with (often the same) children, we demonstrate that a standard three-level model makes overly restrictive assumptions about the correlation structure, in particular that the within-individual correlation (over time) is greater than the between-partner correlation and that the between-partner correlation is the same regardless of whether partner reports are at the same or different waves.  We propose an extended random effects model which avoids these unrealistic assumptions and makes use of all available data, rather than collapsing to couple-level outcomes.  Additionally, we define `couple' clusters that accommodate changes in partner over the observation period.  A further methodological contribution is the development of an MCMC algorithm for computationally efficient estimation of multivariate random effects probit models.

In relation to the first and second research questions set out in Section \ref{sec:previous.research}, we find that the provision of practical and financial help to parents by adult children, and vice versa, is strongly patterned by both parental and child socio-economic and socio-demographic characteristics.  Broadly speaking, and consistent with previous evidence on the correlates of intergenerational exchanges, parents with greater needs are more likely to receive help, whilst parents with greater capacity are more likely to give help. The same applies to adult children. In terms of needs, for example, older parents and those with limiting illness are more likely to receive practical help from their children than younger and healthier parents, as are those who live alone (consistent with \cite{grundy.read.2012}). Adult children who themselves have preschool or primary age children are more likely to receive practical help from their parents, a finding that resonates with the increasing recognition of the crucial role played by grandparents in providing childcare \citep{digessa.etal.2000}. Parents, and adult children, with fewer socio-economic resources are generally more likely to receive practical and/or financial help than those who are better off, although many of these associations are quite weak, and some indicators point in the opposite direction: for example, adult children with lower levels of education are less likely to receive practical help from their parents than those with higher levels of education, after taking account of other characteristics. This nuances findings from previous studies, and suggests that it may be important to look at the needs (and capacities) of parents relative to their adult children (and vice versa), rather than thinking about them in absolute terms. 

In terms of capacity, geographical proximity is a key factor, a finding from previous studies \citep[e.g.][]{grundy.shelton.2001,grundy.murphy.2006} strongly enforced here: children who live further from their parents have a substantially lower probability of providing practical help, and vice versa. Distance also has a weak negative association with financial help, despite the fact that there is no logistical impediment to providing financial assistance at a distance. This points towards the possibility that distance is partly endogenous: parents and children who choose to live further apart may in general be less close in emotional terms as well as practical ones.

Gender and ethnicity also play a significant role. These characteristics can be interpreted as reflecting social norms in addition to variations in needs and capacities. Women are more likely to give practical support and men are more likely to give financial support; women -- whether parents or children -- are also more likely than men to receive help of any kind. Asian and Asian British children are more likely than White children to provide financial and/or practical help to their parents but are less likely to receive financial or practical help from their parents, suggesting a strong flow of support up the generations. Black and Black British children are somewhere in between: they are as likely as their Asian and Asian British counterparts to be providing financial help to their parents, but they are less likely to be providing practical help (partly because of their tendency to live further apart), whilst their receipt of help from parents is at a similar rate to their White counterparts. The difference between the generations in poverty rates within Asian families is greater than for White or Black families in Britain \citep{RDU.2021}, so this finding is consistent with the idea that support may flow more strongly from those with greater relative capacity to those with greater relative need, but it could also reflect differences in cultural norms \citep{willis.2012}. These findings on ethnic variations in intergenerational support make an important contribution to the evidence base for the UK, which has been hitherto limited, but clearly more in-depth, possibly qualitative, research is needed to investigate the underlying mechanisms.

In relation to the third research question on reciprocity, we find that there are positive adjusted correlations between giving and receiving practical help, and low -- but not negative -- correlations between giving and receiving financial help, whether these are assessed from a parental or child perspective. This finding supports an interpretation of moderately strong mutuality between parents and children at a point in time. Moreover, in relation to question 4, on substitution, we find that the cross-correlations between financial and practical help (giving or receiving) are all positive, after adjusting for covariates, suggesting complementarity between the two types of help rather than a tendency towards substitution of financial for practical help or vice versa.

Taken together, our findings point towards considerable heterogeneity in the intergenerational support being provided by, and available to, individuals and families in different circumstances, depending not only on the needs and capacities of the respective parties but also on cultural and family norms. To characterise the pattern as a net upwards or downwards transfer is an over-simplification and there is evidence of a substantial degree of mutuality between parents and children. However, the flipside of that is that parents, and children, in families for whom intergenerational exchange is not possible or not the norm are deprived of a potentially vital system of support that runs in parallel to public welfare. The prevalence and distribution of these private transfers of money and time, as noted in Section \ref{sec:intro}, have substantial implications for well-being across the life course.

The focus of our study is inter-household exchanges of support.  We exclude coresident parents and adult children because data on their exchanges were not collected, but consequently we ignore one form of (mutual) support, especially in Asian and Asian British families among whom the prevalence of multigenerational living is particularly high \citep{nafilyan.etal.2021}; this is an important complement to our finding of high levels of support among Asian and Asian British children for their non-coresident parents.  We also carry out separate analyses of respondent-parent and respondent-child exchanges.  For respondents with non-coresident parents and children (who contribute to both analyses), joint analysis of their exchanges with each generation could answer questions such as the extent to which individuals with a high propensity to support their children also tend to support their parents, the so-called `sandwich' generation \citep{grundy.henretta.2006}. Another limitation of our study is that the relatively short observation window does not permit analysis of reciprocity of exchanges over the lifecourse. The results from our analysis of concurrent reciprocity using panel data should be considered alongside previous research using birth cohort data \citep[e.g.][]{evandrou.etal.2018} which include measures of parental support given to cohort members during childhood.  As noted in Section \ref{sec:data.UKHLS}, a limitation shared by other nationally-representative surveys is that data on support given and received is collected from only one member of the child-parent dyad and little is known about the respondent's non-coresident relative. If there is a tendency for respondents to overstate the help they give and understate the help received, the mutuality of exchanges will be underestimated.

\section*{Acknowledgements} This research was supported by a UK Economic and Social Research Council (ESRC) grant ``Methods for the Analysis of Longitudinal Dyadic Data with an Application to Inter-generational Exchanges of Family Support'' (ref. ES/P000118/1). Additional funding for EG was provided by the Economic and Social Research Council (UK) Research Centre on Micro- Social Change at the University of Essex (grant number ES/L009153/1). Additional funding for SZ was provided by Shanghai Science and Technology Committee Rising-Star Program (22YF1411100).

\appendix

\setcounter{equation}{0}
\setcounter{table}{0}
\renewcommand{\theequation}{S\arabic{equation}}
\renewcommand{\thesection}{Appendix \Alph{section}}
\renewcommand{\thetable}{S\arabic{table}}

\section{MCMC estimation of multivariate random effects probit models}
\label{sec.sup:estimation}

The estimation algorithm described below is implemented in the open source R package \texttt{mvreprobit} \citep{mvreprobit}.  The package and R scripts to carry out the analysis on synthetic data based on the UKHLS data and some other examples can be found at \url{https://github.com/slzhang-fd/mvreprobit}.

\subsection{The models}
\label{sec.sup:estimation.models}

\paragraph{The model for respondent-parent exchanges}
~
For each binary response $y_{rti}^{(C)}$, there exists an underlying continuous variable $y_{rti}^{(C)*}$ such that
\begin{equation}
    y_{rti}^{(C)} = 1 \Leftrightarrow y_{rti}^{(C)*} > 0,
\end{equation}
%with response probability specified by $p(y_{rti}^{(C)}=1) = p(y_{rti}^{(C)*}>0)$,
where $r=1,\dots,4,t=1,\dots,T_i^{(C)},i=1,...,n^{(C)}$. The underlying variables $\boldsymbol{y}_{ti}^{(C)*} = (y_{1ti}^{(C)*},\cdots,y_{4ti}^{(C)*})^\top$ satisfy
\begin{equation}
\label{eq:RPmodel1}
  \boldsymbol{y}_{ti}^{(C)*} = \mathbf{B}^{(C)} \boldsymbol x_{ti}^{(C)} + \boldsymbol{u}_{i}^{(C)} + \boldsymbol{e}_{ti}^{(C)},
\end{equation}
where $\mathbf{B}^{(C)} = (\boldsymbol\beta_1^{(C)},\cdots,\boldsymbol\beta_4^{(C)})^\top$ are coefficient parameters. Assume individual random effect $\boldsymbol{u}_i^{(C)}\sim N(\boldsymbol{0},\mathbf\Sigma_u^{(C)})$, and time-varying residual $\boldsymbol{e}_{ti}^{(C)}\sim N(\boldsymbol{0},\mathbf\Sigma_e^{(C)}).$ For identification, the diagonal elements of $\mathbf\Sigma_e^{(C)}$ are set to one (i.e., $\mathbf\Sigma_e^{(C)}$ is a correlation matrix).

\paragraph{The model for respondent-child exchanges}

Similarly, we specify an extended multivariate random effects probit model for respondent-child exchanges to account for correlated reports from mother and father respondents living in the same household. That is, for $t=1,\dots,T_{ij}^{(P)},i=1,...,n_j^{(P)},j=1,...,n^{(P)},$
\begin{equation}
\label{eq:RCmodel1}
  \boldsymbol{y}_{tij}^{(P)*} = \mathbf{B}^{(P)} \boldsymbol x_{tij}^{(P)} + \boldsymbol{u}_{ij}^{(P)}+ \boldsymbol{v}_{j}^{(P)}+ \boldsymbol{w}_{tj}^{(P)} + \boldsymbol{e}_{tij}^{(P)},
\end{equation}
where underlying response variables $\boldsymbol{y}_{tij}^{(P)*} = (y_{1tij}^{(P)*},\cdots,y_{4tij}^{(P)*})^\top$, $\mathbf{B}^{(P)} = (\boldsymbol\beta_1^{(P)},\cdots,\boldsymbol\beta_4^{(P)})^\top$ are coefficient parameters. Assume individual-specific random effect $\boldsymbol{u}_{ij}^{(P)}\sim N(\boldsymbol{0},\mathbf\Sigma_u^{(P)})$, time-invariant couple-level random effect $\boldsymbol{v}_{j}^{(P)}\sim N(\boldsymbol{0},\mathbf\Sigma_v^{(P)})$, time-varying couple-level random effect $\boldsymbol{w}_{tj}^{(P)}\sim N(\boldsymbol{0},\mathbf\Sigma_w^{(P)})$, and time-varying residual $\boldsymbol{e}_{tij}^{(P)}\sim N(\boldsymbol{0},\mathbf\Sigma_e^{(P)})$.
For identifiability, the diagonal elements of $\mathbf\Sigma_e^{(P)}$ are set to one.

\subsection{Estimation algorithm}
\label{sec.sup:estimation.algorithm}

We propose an efficient Gibbs sampling procedure for a class of multivariate random effect probit models, including the models of \cref{eq:RPmodel1} and \cref{eq:RCmodel1} as special cases.

For simplicity, we omit the superscripts $C$ ad $P$  without ambiguity. For a correlation matrix $\mathbf\Sigma_e$ of dimension $K$ ($K=4$ in our case), let $\boldsymbol\rho_e$ be a length $K(K-1)/2$ vector containing all the correlation coefficients. Note that the $\boldsymbol\rho_e$ of all positive semi-definite correlation matrices form a convex solid body $C_\rho\in [-1,1]^L, L=K(K-1)/2$ \citep{rousseeuw.molenberghs.1994}.
Denote the model parameters by $\boldsymbol\psi=(\boldsymbol{\psi}_\beta, \boldsymbol\psi_\sigma)$, where $\boldsymbol{\psi}_\beta=(\boldsymbol{\beta}_1,\boldsymbol{\beta}_2,\boldsymbol{\beta}_3,\boldsymbol{\beta}_4)$ contains the coefficients and $\boldsymbol\psi_\sigma$ contains the variance-covariance parameters of latent variables.

For the respondent-parent exchanges model in \cref{eq:RPmodel1}, we have $\boldsymbol\psi_\sigma = (\mathbf\Sigma_u,\boldsymbol\rho_e)$ and the likelihood function is
\begin{equation}
\label{eq:likelihood_eq1}
  \begin{aligned}
    p(\boldsymbol y|\mathbf X,\boldsymbol\psi)&=\int p(\boldsymbol y|\mathbf X,\boldsymbol u,\boldsymbol\psi_\beta,\boldsymbol{\rho}_e) p(\boldsymbol u|\mathbf\Sigma_u)d\boldsymbol u \\
    &= \prod_{i=1}^n\int\left[\prod_{t=1}^{T_i}\int_{A_{ti}}\phi(\boldsymbol y^*_{ti}|\boldsymbol x_{ti},\boldsymbol u_{i},\boldsymbol\psi_\beta,\boldsymbol{\rho}_e)d\boldsymbol y_{ti}^*\right]\phi(\boldsymbol u_i|\mathbf\Sigma_u)d\boldsymbol u_i,
  \end{aligned}
\end{equation}
where $\phi$ is the multivariate normal probability density function, $A_{ti}=\bigtimes_{r=1}^4 (a_{rti},b_{rti})$, $\bigtimes$ is the Cartesian product, and
\begin{equation}
\label{eq:intervals_eq1}
  (a_{rti},b_{rti})=
  \begin{cases}
    (-\infty, 0),& \text{if } y_{rti}=0,\\
    (0,  \infty), & \text{otherwise}.
  \end{cases}
\end{equation}
% $A_{rti} = (-\infty,0)$ if $y_{rti}=0$ and $A_{rti} =(0,\infty)$ if $y_{rti}=1$.

Similarly, for the respondent-child exchanges model in \cref{eq:RCmodel1}, we have $\boldsymbol\psi_\sigma = (\mathbf\Sigma_u,\mathbf\Sigma_w,\mathbf\Sigma_v,\boldsymbol\rho_e)$ and the likelihood function is
\begin{equation}\label{eq:likelihood_eq2}
  \begin{aligned}
    p(\boldsymbol y|\mathbf X,\boldsymbol\psi) &=\int\int\int p(\boldsymbol y|\mathbf X,\boldsymbol{w},\boldsymbol{u},\boldsymbol{v},\boldsymbol\psi_\beta,\boldsymbol{\rho}_e)p(\boldsymbol{w}|\mathbf\Sigma_w)p(\boldsymbol{u}|\mathbf\Sigma_u)p(\boldsymbol{v}|\mathbf\Sigma_v)d\boldsymbol{w}d\boldsymbol{u}d\boldsymbol{v}\\
    &=\prod_{j=1}^n\int\Bigg\{\prod_{i=1}^{n_j}\int\Bigg[\prod_{t=1}^{T_{ij}}\int\int_{A_{tij}}\phi(\boldsymbol y^*_{tij}|\boldsymbol x_{tij},\boldsymbol u_{ij},\boldsymbol v_{j},\boldsymbol w_{tj},\boldsymbol\psi_\beta,\boldsymbol\rho_e)d\boldsymbol y_{tij}^*\\
    &\quad\times \phi(\boldsymbol w_{tj}|\mathbf\Sigma_w)d\boldsymbol{w}_{tj}\Bigg] \phi(\boldsymbol u_{ij}|\mathbf\Sigma_u)d\boldsymbol u_{ij}\Bigg\} \phi(\boldsymbol v_j|\mathbf\Sigma_v)d\boldsymbol{v}_j,
  \end{aligned}
\end{equation}
where $A_{tij}=\bigtimes_{r=1}^4 (a_{rtij},b_{rtij})$ and
\begin{equation}\label{eq:intervals_eq2}
  (a_{rtij},b_{rtij})=
  \begin{cases}
    (-\infty, 0),& \text{if } y_{rtij}=0,\\
    (0,  \infty), & \text{otherwise}.
  \end{cases}
\end{equation}
% $(a_{rtij},b_{rtij}) = (-\infty,0)$ if $y_{rtij}=0$ and $(a_{rtij},b_{rtij}) =(0,\infty)$ if $y_{rtij}=1$.

The above likelihood functions involve complex integrals that make estimation by maximum likelihood impractical. We instead propose an efficient Gibbs procedure to estimate model parameters $\boldsymbol\psi$. An overview of the sampling steps is given in the rest of this section, with further detail provided in Section \ref{sec.sup:estimation.details}.

\begin{itemize}
  \item \textbf{Sampling of latent variables.}
  Denote all latent variables by $\boldsymbol\zeta=(\boldsymbol{y}^*,\boldsymbol u,\boldsymbol v,\boldsymbol w)$.
  Given the model parameters $\boldsymbol{\psi}$, sample each element of $\boldsymbol{\zeta}$.
  First, sample $\boldsymbol y^*$ given the observed response $\boldsymbol y$, parameters $\boldsymbol\psi$ and remaining latent variables $(\boldsymbol w,\boldsymbol u,\boldsymbol v)$.  Writing
  \begin{equation}
    \boldsymbol y_{tij}^{*} = \boldsymbol\mu_{tij} + \boldsymbol e_{tij},
  \end{equation}
  where $\boldsymbol\mu_{tij}= \boldsymbol\psi_\beta \boldsymbol x_{tij} + \boldsymbol u_{ij}+ \boldsymbol v_{j}+ \boldsymbol w_{tj}$, we sample $\boldsymbol y_{tij}^*$ from a truncated multivariate normal distribution with mean $\boldsymbol{\mu}_{tij}$ and covariance matrix $\mathbf\Sigma_e$ on $A_{tij}$ using the Cholesky factorization. This sampling method is similar to the GHK algorithm \citep{geweke.1991,hajivassiliou.etal.1996,keane.1993}.

  We next sample $\boldsymbol u$, given $\boldsymbol\psi$ and $(\boldsymbol y^*,\boldsymbol w,\boldsymbol v)$ from the posterior
  \begin{equation}
    p(\boldsymbol u|\boldsymbol y^*,\boldsymbol w,\boldsymbol v ,\boldsymbol\psi)\propto p(\boldsymbol y^*|\boldsymbol u,\boldsymbol w,\boldsymbol v,\boldsymbol\psi_\beta,\boldsymbol\rho_e) p(\boldsymbol{u}|\mathbf\Sigma_u).
  \end{equation}

  The sampling steps for the other random effects $(\boldsymbol{w},\boldsymbol{v})$ are essentially the same as sampling $\boldsymbol{u}$.

  \item \textbf{Sampling of parameters.}
  Given sampled values of the latent variables $\boldsymbol\zeta$, we then sample the parameters $\boldsymbol{\psi}=(\boldsymbol{\psi}_\beta,\boldsymbol{\psi}_\sigma),$ where $\boldsymbol{\psi}_\sigma = (\mathbf\Sigma_u,\mathbf\Sigma_v,\mathbf\Sigma_w,\bm \rho_e)$.

  To sample the coefficients $\boldsymbol\psi_\beta = (\boldsymbol{\beta}_1,\boldsymbol{\beta}_2,\boldsymbol{\beta}_3,\boldsymbol{\beta}_4)$, given $\boldsymbol{\psi}_\sigma$ and $\boldsymbol{\zeta}$, we draw from the posterior
  \begin{equation}
    p(\boldsymbol{\psi}_\beta|\boldsymbol{y}^*,\boldsymbol{u},\boldsymbol{v},\boldsymbol{w},\boldsymbol{\psi}_\sigma)\propto p(\boldsymbol{y}^*|\boldsymbol{u},\boldsymbol{v},\boldsymbol{w},\boldsymbol{\psi}_\beta,\boldsymbol{\psi}_\sigma) p(\boldsymbol{\psi}_\beta),
  \end{equation}
  where the weak-information prior distribution $p(\boldsymbol{\psi}_\beta)$ is set to be independently normal with mean 0 and variance 100.

  The variance-covariance matrix $\mathbf\Sigma_u$ is sampled from the posterior distribution
  \begin{equation}
    p(\mathbf\Sigma_u|\boldsymbol u)\propto p(\boldsymbol{u}|\mathbf \Sigma_u)p(\mathbf\Sigma_u),
  \end{equation}
  where the conjugate inverse Wishart prior $p(\mathbf\Sigma_u) = \mathcal{IW}(\mathbf I,4)$ is used, $\mathbf I$ is the identity matrix. We sample $\mathbf\Sigma_v$ and $\mathbf\Sigma_w$ in a similar way.

 Finally, in order to sample $\bm \rho_e$, a positive semi-definite constraint is required to ensure that $\mathbf\Sigma_e$ is a proper correlation matrix.  A random-walk Metropolis sampler with joint uniform prior $p(\boldsymbol{\rho}_e) = I(\boldsymbol{\rho}_e\in C_\rho)$ is used. To determine whether a proposal $\boldsymbol{\rho}_e'$ is in $C_\rho$, we use a proposition of Sylvester’s criterion (see Section \ref{sec.sup:estimation.details} for details). Let $\mathbf\Sigma=(\mathbf\Sigma_u,\mathbf\Sigma_v,\mathbf\Sigma_w)$, the random-walk Metropolis sampling algorithm is given in below.

  \begin{algorithm}[Random-walk Metropolis sampler for $\boldsymbol{\rho}_e$]~

    For $l=1,\dots,6$:
    \begin{enumerate}
      \item Generate proposal
      \begin{equation*}
        \boldsymbol{\rho}_e' = \boldsymbol{\rho}_e+\gamma\boldsymbol{\epsilon}_l,
      \end{equation*}
      where $\boldsymbol{\epsilon}_l$ is a vector with the $l$-th element sampled from the standard normal distribution and the remaining elements are set to 0.
      \item Accept $\boldsymbol{\rho}'_e$ with the probability
      \begin{equation}
        \alpha(\boldsymbol{\rho}_e\rightarrow\boldsymbol{\rho}_e') = \min\left\{1,\frac{p(\boldsymbol{y}^*|\boldsymbol{\zeta},\boldsymbol{\psi}_\beta,\mathbf\Sigma,\boldsymbol{\rho}_e')p(\boldsymbol{\rho}_e')}{p(\boldsymbol{y}^*|\boldsymbol{\zeta},\boldsymbol{\psi}_\beta,\mathbf\Sigma,\boldsymbol{\rho}_e)p(\boldsymbol{\rho}_e)}\right\},
      \end{equation}
      and stay at $\boldsymbol{\rho}_e$ with probability $1-\alpha(\boldsymbol{\rho}_e\rightarrow\boldsymbol{\rho}_e')$.
    \end{enumerate}
  \end{algorithm}
  Note that the step size $\gamma$ is chosen to control rejection rates between 0.7-0.8.
\end{itemize}

% kronecker(Sigma_inv, t(X)) %*% as.vector(Z)
%\appendix
%\setcounter{equation}{0}
%\section{Appendix: Details of the MCMC algorithm}
%\label{app:details_mcmc}

\subsection{Further details of the MCMC algorithm}
\label{sec.sup:estimation.details}

\paragraph{Sampling of latent variable $\boldsymbol y^*$.}

In order to sample $\boldsymbol y^*$, we sample each $\boldsymbol y_{tij}^*$ from the truncated multivariate normal distribution
\begin{equation}
  TN\left(\boldsymbol\mu_{tij}, \mathbf\Sigma_e, \boldsymbol a_{tij},\boldsymbol b_{tij}\right),
\end{equation}
where $\boldsymbol\mu_{tij}=\boldsymbol\psi_\beta \boldsymbol x_{tij} + \boldsymbol u_{ij}+ \boldsymbol v_{j}+ \boldsymbol w_{tj}$ and $(\boldsymbol a_{tij},\boldsymbol b_{tij})$ are interval boundaries defined in \cref{eq:intervals_eq2}.

The Cholesky decomposition of $\mathbf\Sigma_e$ is $\mathbf\Sigma_e = \Gamma_e\Gamma_e^\top$. Since
\begin{equation}\label{eq:recover_y_star}
  \boldsymbol y_{tij}^{*} = \boldsymbol{\mu}_{tij} + \Gamma_e\boldsymbol \epsilon_{tij},
\end{equation}
where each element of $\boldsymbol\epsilon_{tij}$ independently follows a univariate truncated normal distribution, it is equivalent to sample $\boldsymbol\epsilon_{tij}$ from
\begin{equation}
  TN(\boldsymbol{0},I,\Gamma_e^{-1}(\boldsymbol a_{tij}-\boldsymbol{\mu}_{tij}),\Gamma_e^{-1}(\boldsymbol b_{tij}-\boldsymbol{\mu}_{tij})),
\end{equation}
and then transform back to $\boldsymbol y_{tij}^*$ using \cref{eq:recover_y_star}. Note that $\Gamma_e^{-1}$ is also lower triangular and can be computed iteratively by row.

\paragraph{Sampling of latent variable $\boldsymbol{u}$.}
From the normality assumption, we have
\begin{equation}
  \boldsymbol{u}_{ij}|\boldsymbol y_{tij}^*,\boldsymbol{w}_{tj},\boldsymbol{v}_j,\boldsymbol\psi \sim N(\boldsymbol\mu_{u_{ij}},\mathbf\Sigma_{u_{ij}}),
\end{equation}
where
\begin{align*}
  \mathbf\Sigma_{u_{ij}} &= \left( T_{ij}\mathbf\Sigma_e^{-1}+ \mathbf\Sigma_u^{-1} \right)^{-1} \quad\text{and}\\
  \boldsymbol\mu_{u_{ij}} &= \mathbf\Sigma_{u_{ij}}\mathbf\Sigma_e^{-1}\left[\sum_{t=1}^{T_{ij}}(\boldsymbol y^*_{tij} - \boldsymbol v_j-\boldsymbol w_{tj})\right].
\end{align*}

\paragraph{Sampling of coefficients $\boldsymbol\psi_\beta$.}
Assuming normality, we have
\begin{equation}
  \text{vec}(\boldsymbol{\psi}_\beta) |\boldsymbol{y}^*,\boldsymbol{u},\boldsymbol{v},\boldsymbol{w},\boldsymbol{\psi}_\sigma \sim N(\boldsymbol\mu_\beta, \mathbf\Sigma_\beta),
\end{equation}
with
\begin{align*}
  \mathbf\Sigma_\beta &= \left(\mathbf I/\sigma_0^2 + \mathbf\Sigma_e^{-1}\otimes (\mathbf X^\top \mathbf X)\right)^{-1} \quad\text{and}\\
  \boldsymbol\mu_\beta &= \mathbf\Sigma_\beta\left(\mathbf\Sigma_e^{-1}\otimes \mathbf X^\top \right)\text{vec}(\boldsymbol{y}^* - \boldsymbol{u}-\boldsymbol{v}-\boldsymbol{w}),
\end{align*}
where $\otimes$ denotes the Kronecker product, and $\mathbf I$ denotes the identity matrix.

\paragraph{Sampling of $\mathbf\Sigma_u$.}
Due to the conjugate inverse Wishart prior, we have the posterior
\begin{equation}
  \mathbf \Sigma_u|\boldsymbol{u}\sim \mathcal{W}^{-1}(\mathbf I + \mathbf X^\top\mathbf X, 4+n).
\end{equation}

\paragraph{Checking whether $\boldsymbol{\rho}_e'\in C_\rho$.}

We describe the main idea here and refer to \cite{barnard.etal.2000} for further discussion.

We begin with a correlation matrix $\mathbf\Sigma(\boldsymbol{\rho})$ with lower triangular elements $\boldsymbol{\rho}=(\rho_1,\dots\rho_L)$. For a correlation matrix of dimension $K$, without loss of generality, assume $\rho_l$ is in the $K$th row, or we can always swap both row and column without changing the positive definiteness and determinant value. According to Sylvester's criterion, $\mathbf\Sigma(\rho_l',\boldsymbol{\rho}_{-l})$ is positive semi-definite if and only if all leading principal minors $|\mathbf\Sigma_k|,k=1,\dots,K$ are non-negative, where $|\cdot|$ is the matrix determinant, and $\mathbf\Sigma_k$ is the $k$th primary submatrix of $\mathbf\Sigma(\rho_l',\boldsymbol{\rho}_{-l})$. Since $\mathbf\Sigma(\rho_l,\boldsymbol{\rho}_{-l})$ is positive semi-definite, $\vert \mathbf\Sigma_k\vert\geq 0, \text{ for } k=1,\ldots,K-1$.
Thus, $\mathbf\Sigma(\rho_l',\boldsymbol{\rho}_{-l})$ is positive semi-definite if and only if $|\mathbf\Sigma(\rho_l',\boldsymbol{\rho}_{-l})| \geq 0.$

\section{Additional results}

\begin{table}[!htbp]
\caption{\label{tab:corr.RP} Estimates of correlations between pairs of outcomes ($r$ and $r^\prime$) from multivariate random effects probit model for exchanges between non-coresident adult children and parents from a child perspective.  The correlations shown are between time-varying residuals, between individual random effects, and between the total residual (overall). \vspace{4pt}}
\begin{footnotesize}
\begin{center}
%\centering
\begin{tabular}{l d{2.3}@{}l d{2.3}@{}l d{2.3}@{}l }
\hline
Residual: $\mbox{cor}(e_{rti}^{(C)},e_{r^\prime ti}^{(C)})$ & & & & & & \\ [2pt]
Individual: $\mbox{cor}(u_{ri}^{(C)},u_{r^\prime i}^{(C)})$ &   \multicolumn{2}{c}{GP$^a$} & \multicolumn{2}{c}{GF} & \multicolumn{2}{c}{RP} \\
Overall &   \multicolumn{2}{c}{$(r=1)$} & \multicolumn{2}{c}{$(r=2)$} & \multicolumn{2}{c}{$(r=3)$} \\
\hline
%template
%GP & .000 & $^*$ & .000 & $^*$ & .000 & $^*$ & .000 & $^*$ \\
%GP &      & 1    &      &      &      &      &      &    \\
%   &      &      &      &      &      &      &      &    \\
%   &      &      &      &      &      &      &      &    \\ [6pt]
GF        & .465 & $^*$ &      &      &      &         \\
$(r=2)$   & .493 & $^*$ &      &      &      &         \\
          & .482 & $^*$ &      &      &      &         \\ [6pt]
RP        & .490 & $^*$ & .229 & $^*$ &      &         \\
$(r=3)$   & .385 & $^*$ & .072 & $^*$ &      &         \\
          & .430 & $^*$ & .135 & $^*$ &      &         \\ [6pt]
RF        & .304 & $^*$ & .191 &      & .401 & $^*$   \\
$(r=4)$   & .252 & $^*$ & .028 &      & .440 & $^*$    \\
          & .274 & $^*$ & .057 &      & .422 & $^*$    \\
\hline
%templste
%& -0.000 & $^*$ & (0.000) & -0.000 & $^*$ & (0.000) & -0.000 & $^*$ & (0.000) & -0.000 & $^*$ & (0.000) \\
\multicolumn{7}{p{0.55\textwidth}}{$^*$95\% credible interval excludes zero; $^a$GP=give practical help, GF=give financial help, RP=receive practical help, RF=receive financial help.}
\end{tabular}
\end{center}
\end{footnotesize}
\end{table}

%%%%%%%%%%%%%%%%%%%%%%%%%%%%%%%%%%%%%%%%%%%%%%%%%%%%%%%%%%%

\begin{table}[!htbp]
\caption{\label{tab:corr.RC} Estimates of correlations between pairs of outcomes ($r$ and $r^\prime$) from multivariate random effects probit model for exchanges between non-coresident adult children and parents from a parental perspective.  The correlations shown are between time-varying residuals, between each of the three random effects, and between the total residual (overall). Correlations with RF are not estimated because the model for receipt of financial help from children was estimated separately from the model for exchanges of other type of support between respondents and children. \vspace{4pt}}
\begin{footnotesize}
\begin{center}
%\centering
\begin{tabular}{l d{2.3}@{}l d{2.3}@{}l }
\hline
Residual: $\mbox{cor}(e_{rtij},e_{r^\prime tij})$ & & & &  \\ [2pt]
Individual: $\mbox{cor}(u_{rij},u_{r^\prime ij})$ & & & &  \\ [2pt]
Couple (time-varying): $\mbox{cor}(w_{rtj},w_{r^\prime tj})$ & & & & \\ [2pt]
Couple (fixed): $\mbox{cor}(v_{rj},v_{r^\prime j})$ &   \multicolumn{2}{c}{GP$^a$} & \multicolumn{2}{c}{GF}   \\
Overall &   \multicolumn{2}{c}{$(r=1)$} & \multicolumn{2}{c}{$(r=2)$}   \\
\hline
%templste
GF       & .485 & $^*$ &      &               \\
$(r=2)$  & .510 & $^*$ &      &               \\
         & .430 & $^*$ &      &               \\
         & .378 & $^*$ &      &               \\
         & .439 & $^*$ &      &               \\ [6pt]
RP       & .515 & $^*$ & .319 & $^*$          \\
$(r=3)$  & .478 & $^*$ & .319 & $^*$          \\
         & .425 & $^*$ & .338 & $^*$          \\
         & .298 & $^*$ & .112 & $^*$          \\
         & .416 & $^*$ & .260 & $^*$          \\
\hline
%templste
%& -0.000 & $^*$ & (0.000) & -0.000 & $^*$ & (0.000) & -0.000 & $^*$ & (0.000) & -0.000 & $^*$ & (0.000) \\
\multicolumn{5}{p{0.6\textwidth}}{$^*$95\% credible interval excludes zero; $^a$GP=give practical help, GF=give financial help, RP=receive practical help.}
\end{tabular}
\end{center}
\end{footnotesize}
\end{table}

%%%%%%%%%%%%%%%%%%%%%%%%%%%%%%%%%%%%%%%%%%%%%%%%%%%%%%%%%%%

\begin{table}[!htbp]
\caption{\label{tab:modelRP.nodist.coeff} Results from multivariate random effects probit model for exchanges with non-coresident parents from an adult child perspective. Model excluding geographical proximity. The estimates are posterior means from MCMC samples (and posterior standard deviations in parentheses). \vspace{4pt}}
\begin{footnotesize}
\centering
\begin{tabular}{l d{2.3}@{}l r d{2.3}@{}l r d{2.3}@{}l r d{2.3}@{}l r}
\hline
&   \multicolumn{3}{c}{To parents:} & \multicolumn{3}{c}{To parents:} & \multicolumn{3}{c}{From parents:} & \multicolumn{3}{c}{From parents:} \\
&   \multicolumn{3}{c}{practical} & \multicolumn{3}{c}{financial} & \multicolumn{3}{c}{practical} & \multicolumn{3}{c}{financial} \\
&   \multicolumn{3}{c}{$(r=1)$} & \multicolumn{3}{c}{$(r=2)$} & \multicolumn{3}{c}{$(r=3)$} & \multicolumn{3}{c}{$(r=4)$} \\
\cmidrule(lr){2-4} \cmidrule(lr){5-7} \cmidrule(lr){8-10} \cmidrule(lr){11-13}
Variable & \multicolumn{2}{c}{Est.} & \multicolumn{1}{c}{(SD)} & \multicolumn{2}{c}{Est.} & \multicolumn{1}{c}{(SD)} & \multicolumn{2}{c}{Est.} & \multicolumn{1}{c}{(SD)} & \multicolumn{2}{c}{Est.} & \multicolumn{1}{c}{(SD)} \\
\hline
\multicolumn{13}{l}{\emph{Coefficients of explanatory variables, $\boldsymbol\beta_r^{(C)}$}} \\
%templste
%& -0.000 & $^*$ & (0.000) & -0.000 & $^*$ & (0.000) & -0.000 & $^*$ & (0.000) & -0.000 & $^*$ & (0.000) \\
Age (years)$^a$
& -0.014 & $^*$ & (0.002) &  0.006 &      & (0.003) & -0.052 & $^*$ & (0.002) & -0.058 & $^*$ & (0.003) \\
Age-squared $\times 10^{-1}$
& -0.001 &      & (0.001) & -0.002 &      & (0.001) & -0.002 &      & (0.001) &  0.000 &      & (0.001) \\
Female
&  0.221 &      & (0.024) & -0.079 &      & (0.034) &  0.334 &      & (0.023) &  0.175 &      & (0.025) \\
\multicolumn{13}{l}{Ethnicity (ref=White)} \\
~~Asian/ Asian British
&  0.629 & $^*$ & (0.049) &  0.966 & $^*$ & (0.062) & -0.303 & $^*$ & (0.046) & -0.213 & $^*$ & (0.052) \\
~~Black/ Black British
&  0.091 &      & (0.074) &  1.140 & $^*$ & (0.080) & -0.138 & $^*$ & (0.070) &  0.051 &      & (0.072) \\
~~Other
&  0.017 &      & (0.081) &  0.530 & $^*$ & (0.099) & -0.157 & $^*$ & (0.076) & -0.152 &      & (0.083) \\
Coresident partner
& -0.075 & $^*$ & (0.026) &  0.015 &      & (0.039) & -0.486 & $^*$ & (0.024) & -0.420 & $^*$ & (0.027) \\
Long-term illness
& -0.107 & $^*$ & (0.027) &  0.002 &      & (0.042) &  0.048 &      & (0.027) &  0.138 & $^*$ & (0.031) \\
Post-school education
& -0.340 & $^*$ & (0.024) &  0.120 & $^*$ & (0.033) & -0.143 & $^*$ & (0.023) & -0.075 & $^*$ & (0.026) \\
Unemp./ econ. inactive
& -0.025 &      & (0.025) & -0.070 &      & (0.038) & -0.083 & $^*$ & (0.024) &  0.107 & $^*$ & (0.028) \\
Log annual hh inc.
& -0.051 & $^*$ & (0.012) &  0.104 & $^*$ & (0.020) & -0.000 &      & (0.012) & -0.144 & $^*$ & (0.012) \\
Home owner
&  0.115 & $^*$ & (0.027) & -0.077 & $^*$ & (0.039) &  0.232 & $^*$ & (0.025) & -0.310 & $^*$ & (0.026) \\
\multicolumn{13}{l}{Child coresidence status (ref=none)}   \\
~~Cores. only$^b$
& -0.007 &      & (0.035) & -0.022 &      & (0.057) &  0.866 & $^*$ & (0.033) &  0.061 & $^*$ & (0.037) \\
~~Cores. and non-cores$^b$
&  0.052 &      & (0.044) & -0.028 &      & (0.071) &  0.757 & $^*$ & (0.043) &  0.141 & $^*$ & (0.050) \\
~~Non-cores. only
&  0.220 & $^*$ & (0.039) & -0.107 &      & (0.057) &  0.140 & $^*$ & (0.042) &  0.077 & $^*$ & (0.048) \\
\multicolumn{13}{l}{Age of youngest coresident child (ref=$<2$ yrs)}  \\
~~2--4 yrs
&  0.074 & $^*$ & (0.032) & -0.000 &      & (0.055) &  0.041 &      & (0.031) &  0.052 &      & (0.035) \\
~~5--10 yrs
&  0.089 & $^*$ & (0.035) & -0.010 &      & (0.059) &  0.013 &      & (0.033) &  0.096 & $^*$ & (0.039) \\
~~11--16 yrs
&  0.144 & $^*$ & (0.040) & -0.053 &      & (0.066) & -0.426 & $^*$ & (0.039) &  0.051 &      & (0.047) \\
~~$>$ 16 yrs
&  0.220 & $^*$ & (0.046) & -0.109 &      & (0.073) & -0.747 & $^*$ & (0.047) &  0.034 & $^*$ & (0.055) \\
\multicolumn{13}{l}{Number of siblings (ref=none)} \\
~~1
& -0.087 & $^*$ & (0.035) & -0.156 & $^*$ & (0.053) &  0.108 & $^*$ & (0.036) & -0.082 & $^*$ & (0.039) \\
~~$\geq$ 2
& -0.086 & $^*$ & (0.034) & -0.057 &      & (0.048) & -0.126 & $^*$ & (0.035) & -0.264 & $^*$ & (0.037) \\
Age of oldest parent (yrs)$^a$
&  0.034 & $^*$ & (0.002) &  0.008 & $^*$ & (0.003) & -0.016 & $^*$ & (0.002) &  0.020 & $^*$ & (0.002) \\
Age-squared $\times 10^{-1}$
&  0.009 & $^*$ & (0.001) &  0.011 & $^*$ & (0.001) & -0.011 & $^*$ & (0.001) & -0.002 &      & (0.001) \\
$\geq$ 1 parent lives alone
&  0.670 & $^*$ & (0.021) &  0.508 & $^*$ & (0.031) & -0.124 & $^*$ & (0.021) &  0.016 &      & (0.024) \\
Constant
& -0.096 &      & (0.126) & -3.844 &      & (0.218) & -0.615 &      & (0.131) &  0.296 &      & (0.129) \\
\emph{Random effect variance} $\sigma_{urr}^{(C)}$
&  2.030 &      & (0.054) &  1.813 &      & (0.082) &  1.561 &      & (0.044) &  1.340 &      & (0.049) \\
\emph{Within-individual correlation}
&  0.670 &      & (0.006) &  0.644 &      & (0.010) &  0.609 &      & (0.007) &  0.572 &      & (0.009) \\
\hline
\multicolumn{13}{p{\textwidth}}{$^*$95\% credible interval does not include zero; $^a$Age is centred around 40 and parental age around 70; squared age and parental age are also included in the model, but their estimates are not shown because their effects are negligible. $^b$Contrasts 1+ coresident child where youngest is age $<$2 years versus no children. $^c$Effects of age of youngest child among respondents with coresident children.}
\end{tabular}
\end{footnotesize}
\end{table}

%%%%%%%%%%%%%%%%%%%%%%%%%%%%%%%%%%%%%%%%%%%%%%

\begin{table}[!htbp]
\caption{\label{tab:modelRC.nodist.coeff} Results from multivariate random effects probit model for exchanges with non-coresident children from a parental perspective. Model excluding geographical proximity. The estimates are posterior means from MCMC samples (and posterior standard deviations in parentheses). \vspace{4pt}}
\begin{footnotesize}
\centering
\begin{tabular}{l d{2.3}@{}l r d{2.3}@{}l r d{2.3}@{}l r d{2.3}@{}l r}
\hline
&   \multicolumn{3}{c}{To children:} & \multicolumn{3}{c}{To children:} & \multicolumn{3}{c}{From children:} & \multicolumn{3}{c}{From children:} \\
&   \multicolumn{3}{c}{practical} & \multicolumn{3}{c}{financial} & \multicolumn{3}{c}{practical} & \multicolumn{3}{c}{financial} \\
&   \multicolumn{3}{c}{$(r=1)$} & \multicolumn{3}{c}{$(r=2)$} & \multicolumn{3}{c}{$(r=3)$} & \multicolumn{3}{c}{$(r=4)$} \\
\cmidrule(lr){2-4} \cmidrule(lr){5-7} \cmidrule(lr){8-10} \cmidrule(lr){11-13}
Variable & \multicolumn{2}{c}{Est.} & \multicolumn{1}{c}{(SD)} & \multicolumn{2}{c}{Est.} & \multicolumn{1}{c}{(SD)} & \multicolumn{2}{c}{Est.} & \multicolumn{1}{c}{(SD)} & \multicolumn{2}{c}{Est.} & \multicolumn{1}{c}{(SD)} \\
\hline
\multicolumn{13}{l}{\emph{Coefficients of explanatory variables, $\boldsymbol\beta_r^{(P)}$}} \\
%templste
%& -0.000 & $^*$ & (0.000) & -0.000 & $^*$ & (0.000) & -0.000 & $^*$ & (0.000) & -0.000 & $^*$ & (0.000) \\
Age (years)
&  0.055 & $^*$ & (0.005) & -0.036 & $^*$ & (0.005) &  0.016 & $^*$ & (0.005) & -0.010 &      & (0.010) \\
Age squared $\times 10^{-1}$
& -0.024 & $^*$ & (0.001) & -0.003 & $^*$ & (0.001) &  0.004 & $^*$ & (0.001) &  0.000 & $^*$ & (0.002) \\
Female
&  0.398 & $^*$ & (0.028) & -0.295 & $^*$ & (0.025) &  0.664 & $^*$ & (0.030) &  0.380 & $^*$ & (0.077) \\
\multicolumn{13}{l}{Ethnicity (ref=White)} \\
~~Asian/ Asian British
& -0.763 & $^*$ & (0.095) & -0.944 & $^*$ & (0.096) &  0.347 & $^*$ & (0.088) &  1.652 & $^*$ & (0.242) \\
~~Black/ Black British
& -0.327 & $^*$ & (0.106) & -0.118 &      & (0.098) &  0.047 &      & (0.100) &  1.628 & $^*$ & (0.242) \\
~~Other
& -0.089 &      & (0.138) &  0.032 &      & (0.124) &  0.114 &      & (0.134) &  0.799 & $^*$ & (0.240) \\
Coresident partner
&  0.196 & $^*$ & (0.039) & -0.005 &      & (0.036) & -0.770 & $^*$ & (0.040) & -0.034 &      & (0.123) \\
Long-term illness
& -0.292 & $^*$ & (0.029) &  0.008 &      & (0.028) &  0.383 & $^*$ & (0.028) &  0.408 & $^*$ & (0.079) \\
Post-school education
& -0.032 &      & (0.037) &  0.527 & $^*$ & (0.034) & -0.349 & $^*$ & (0.036) & -0.151 &      & (0.081) \\
Unemp./ econ. inactive
&  0.189 & $^*$ & (0.036) & -0.256 & $^*$ & (0.034) & -0.349 & $^*$ & (0.036) & -0.151 &      & (0.081) \\
Log annual hh inc.
& -0.032 &      & (0.020) &  0.195 & $^*$ & (0.020) & -0.047 & $^*$ & (0.019) & -0.152 & $^*$ & (0.040) \\
Home owner
&  0.372 & $^*$ & (0.045) &  0.430 & $^*$ & (0.042) & -0.257 & $^*$ & (0.042) & -0.730 & $^*$ & (0.120) \\
Any coresident children
& -0.087 & $^*$ & (0.041) & -0.206 & $^*$ & (0.038) & -0.006 &      & (0.040) &  0.266 & $^*$ & (0.086) \\
\multicolumn{13}{l}{No. non-cores. children (ref=1)} \\
~~2
&  0.496 & $^*$ & (0.041) &  0.104 & $^*$ & (0.037) &  0.322 & $^*$ & (0.040) &  0.178 & $^*$ & (0.099) \\
~~$\geq$ 3
&  0.699 & $^*$ & (0.047) &  0.182 & $^*$ & (0.040) &  0.679 & $^*$ & (0.045) &  0.455 & $^*$ & (0.106) \\
Has a surviving parent
& -0.191 & $^*$ & (0.037) & -0.064 &      & (0.034) & -0.339 & $^*$ & (0.038) & -0.456 & $^*$ & (0.101) \\
Constant
& -0.097 &      & (0.207) & -2.087 &      & (0.208) & -0.888 &      & (0.204) & -2.653 &      & (0.493) \\
\multicolumn{13}{l}{\emph{Random effect variances}} \\
Time-invariant couple $\sigma_{vrr}^{(P)}$
&  2.278 &      & (0.119) &  1.440 &      & (0.075) &  1.869 &      & (0.096) &  1.733 &      & (0.361) \\
Time-varying couple $\sigma_{wrr}^{(P)}$
&  1.472 &      & (0.114) &  1.248 &      & (0.093) &  1.343 &      & (0.101) &  2.123 &      & (0.773) \\
Time-invariant ind. $\sigma_{urr}^{(P)}$
&  0.888 &      & (0.078) &  0.703 &      & (0.061) &  0.763 &      & (0.068) &  0.782 &      & (0.404) \\
\multicolumn{13}{l}{\emph{Intra-cluster correlations}} \\
Within-individual
&  0.561 &      & (0.007) &  0.488 &      & (0.008) &  0.529 &      & (0.008) &  0.447 &      & (0.023) \\
Within-couple
&  0.665 &      & (0.009) &  0.612 &      & (0.010) &  0.645 &      & (0.010) &  0.686 &      & (0.033) \\
\hline
\multicolumn{13}{p{\textwidth}}{$^*$95\% credible interval does not include zero.}
\end{tabular}
\end{footnotesize}
\end{table}

%%%%%%%%%%%%%%%%%%%%%%%%%%%%%%%%%%%%%%%

\begin{table}[!htbp]
\caption{\label{tab:modelRP.nodist.prob} Predicted marginal probabilities of giving and receiving practical and financial help to/from parents (from an adult child perspective), calculated from model excluding geographical proximity. \vspace{4pt} }
\begin{footnotesize}
\centering
\begin{tabular}{l cccc}
\hline
Respondent (child) & To parents: & To parents: & From parents: & From parents: \\
characteristics & practical & financial & practical & financial \\
& $(r=1)$ & $(r=2)$ & $(r=3)$ & $(r=4)$ \\
\hline
%template
%&  .000   &   .000    &   .000    &   .000 \\
\multicolumn{5}{l}{Age} \\
~~30 years
&  .468   &   .057    &   .478    &   .242 \\
%~~40
%&  .440   &   .062    &   .364    &   .142 \\
~~50
&  .409   &   .065    &   .252    &   .076 \\
\multicolumn{5}{l}{Gender} \\
~~Male
&  .403   &   .065    &   .315    &   .121 \\
~~Female
&  .450   &   .059    &   .381    &   .144 \\
\multicolumn{5}{l}{Ethnicity} \\
~~Asian or Asian British
&  .558   &   .146    &   .300    &   .110 \\
~~Black or Black British
&  .441   &   .171    &   .331    &   .143 \\
~~White
&  .422   &   .053    &   .358    &   .136 \\
~~Other
&  .426   &   .096    &   .327    &   .117 \\
\multicolumn{5}{l}{Partnership status} \\
~~Unpartnered
&  .442   &   .061    &   .428    &   .178 \\
~~Partner
&  .426   &   .062    &   .331    &   .119 \\
\multicolumn{5}{l}{Has long-term illness} \\
~~No
&  .433   &   .061    &   .352    &   .132 \\
~~Yes
&  .410   &   .062    &   .362    &   .151 \\
\multicolumn{5}{l}{Has post-school education} \\
~~No
&  .463   &   .058    &   .366    &   .139 \\
~~Yes
&  .390   &   .066    &   .338    &   .129 \\
\multicolumn{5}{l}{Employment status} \\
~~Unemployed/econ. inactive
&  .426   &   .058    &   .341    &   .145 \\
~~Employed
&  .431   &   .063    &   .357    &   .131 \\
\multicolumn{5}{l}{Equivalised household income} \\
~~10th percentile
&  .436   &   .057    &   .353    &   .145 \\
%~~25th
%&  .433   &   .059    &   .353    &   .138 \\
~~50th
&  .429   &   .062    &   .353    &   .132 \\
%~~75th
%&  .426   &   .064    &   .353    &   .126 \\
~~90th
&  .377   &   .105    &   .353    &   .061 \\
\multicolumn{5}{l}{Housing tenure} \\
~~Social/private rent or other
&  .413   &   .065    &   .323    &   .161 \\
~~Own home
&  .437   &   .060    &   .368    &   .119 \\
\multicolumn{5}{l}{Presence/age of children} \\
~~None
&  .408   &   .065    &   .268    &   .124 \\
~~Coresident, youngest $<$ 2 yrs
&  .406   &   .064    &   .445    &   .132 \\
~~Coresident, youngest 2--4
&  .422   &   .064    &   .454    &   .138 \\
~~Coresident, youngest 5--10
&  .425   &   .063    &   .448    &   .144 \\
~~Coresident, youngest 11--16
&  .437   &   .060    &   .354    &   .138 \\
~~Coresident, youngest $>$ 16
&  .453   &   .056    &   .290    &   .136 \\
~~Cores. and non-cores, youngest $>$ 16
&  .466   &   .056    &   .270    &   .147 \\
~~Non-coresident only
&  .455   &   .058    &   .294    &   .134 \\
\multicolumn{5}{l}{Number of siblings} \\
~~0
&  .447   &   .067    &   .360    &   .159 \\
~~1
&  .428   &   .057    &   .381    &   .147 \\
~~$\geq$ 2
&  .428   &   .063    &   .335    &   .124 \\
\multicolumn{5}{l}{Age of oldest parent}  \\
%~~60 years
%&  .344   &   .052    &   .386    &   .119 \\
~~70
&  .396   &   .050    &   .373    &   .145 \\
~~80
&  .492   &   .062    &   .318    &   .171 \\
~~90
&  .629   &   .098    &   .229    &   .193 \\
\multicolumn{5}{l}{At least one parent lives alone} \\
~~No
&  .376   &   .048    &   .361    &   .134 \\
~~Yes
&  .523   &   .085    &   .337    &   .136 \\
%\hline
%Overall
%&  .430   &   .062    &   .353    &   .135 \\
\hline
\end{tabular}
\end{footnotesize}
\end{table}

%%%%%%%%%%%%%%%%%%%%%%%%%%%%%%%%%%%%%%%

\begin{table}[!htbp]
\caption{\label{tab:modelRC.nodist.prob} Predicted marginal probabilities of giving and receiving practical and financial help to/from adult children (from a parental perspective), calculated from model excluding geographical proximity. \vspace{4pt} }
\begin{footnotesize}
\centering
\begin{tabular}{l cccc}
\hline
Respondent (parent) & To children: & To children: & From children: & From children: \\
characteristics & practical & financial & practical & financial \\
& $(r=1)$ & $(r=2)$ & $(r=3)$ & $(r=4)$ \\
\hline
%template
%&  .000   &   .000    &   .000    &   .000 \\
\multicolumn{5}{l}{Age} \\
~~50 years
&  .633   &   .401    &   .287    &   .030 \\
%~~60
%&  .604   &   .323    &   .328    &   .027 \\
~~70
&  .495   &   .243    &   .383    &   .025 \\
\multicolumn{5}{l}{Gender} \\
~~Male
&  .480   &   .315    &   .303    &   .021 \\
~~Female
&  .543   &   .270    &   .405    &   .029 \\
\multicolumn{5}{l}{Ethnicity} \\
~~Asian or Asian British
&  .402   &   .167    &   .414    &   .085 \\
~~Black or Black British
&  .470   &   .276    &   .367    &   .084 \\
~~White
&  .522   &   .294    &   .360    &   .021 \\
~~Other
&  .508   &   .299    &   .377    &   .044 \\
\multicolumn{5}{l}{Partnership status} \\
~~Unpartnered
&  .494   &   .290    &   .447    &   .039 \\
~~Partner
&  .525   &   .289    &   .324    &   .018 \\
\multicolumn{5}{l}{Has long-term illness} \\
~~No
&  .531   &   .289    &   .342    &   .022 \\
~~Yes
&  .484   &   .290    &   .402    &   .032 \\
\multicolumn{5}{l}{Has post-school education} \\
~~No
&  .518   &   .264    &   .377    &   .027 \\
~~Yes
&  .513   &   .347    &   .324    &   .023 \\
\multicolumn{5}{l}{Employment status} \\
~~Unemployed/econ. inactive
&  .528   &   .272    &   .369    &   .027 \\
~~Employed
&  .498   &   .312    &   .348    &   .023 \\
\multicolumn{5}{l}{Equivalised household income} \\
~~10th percentile
&  .519   &   .271    &   .366    &   .027 \\
%~~25th
%&  .518   &   .280    &   .363    &   .026 \\
~~50th
&  .516   &   .290    &   .361    &   .025 \\
%~~75th
%&  .514   &   .300    &   .359    &   .024 \\
~~90th
&  .491   &   .450    &   .327    &   .012 \\
\multicolumn{5}{l}{Housing tenure} \\
~~Social/private rent or other
&  .471   &   .239    &   .392    &   .039 \\
~~Own home
&  .530   &   .303    &   .353    &   .020 \\
\multicolumn{5}{l}{Has coresident children} \\
~~No
&  .520   &   .297    &   .362    &   .024 \\
~~Yes
&  .506   &   .266    &   .361    &   .031 \\
{Number of non-coresident children} \\
~~1
&  .447   &   .274    &   .306    &   .021 \\
~~2
&  .525   &   .289    &   .354    &   .024 \\
~~$\geq$ 3
&  .557   &   .301    &   .411    &   .031 \\
\multicolumn{5}{l}{Has surviving parent} \\
~~No
&  .526   &   .292    &   .376    &   .029 \\
~~Yes
&  .496   &   .283    &   .325    &   .019 \\
%\hline
%Overall
%&  .516   &   .289    &   .362    &   .026 \\
\hline
\end{tabular}
\end{footnotesize}
\end{table}

\bibliography{Longitudinal_exchanges_of_support_ref}

\begin{thebibliography}{}

\bibitem[\protect\citeauthoryear{Albertini, Kohli, and Vogel}{Albertini
  et~al.}{2007}]{albertini.etal.2007}
Albertini, M., M.~Kohli, and C.~Vogel (2007).
\newblock Intergenerational transfers of time and money in {E}uropean families:
  common patterns -- different regimes?
\newblock {\em Journal of European Social Policy\/}~{\em 17\/}(4), 319--334.

\bibitem[\protect\citeauthoryear{Attias-Donfut, Ogg, and Wolff}{Attias-Donfut
  et~al.}{2005}]{attias-donfut.etal.2005}
Attias-Donfut, C., J.~Ogg, and F.-C. Wolff (2005).
\newblock European patterns of intergenerational financial and time transfers.
\newblock {\em European Journal of Ageing\/}~{\em 2}, 161--173.

\bibitem[\protect\citeauthoryear{Barnard, McCulloch, and Meng}{Barnard
  et~al.}{2000}]{barnard.etal.2000}
Barnard, J., R.~McCulloch, and X.-L. Meng (2000).
\newblock Modeling covariance matrices in terms of standard deviations and
  correlations, with application to shrinkage.
\newblock {\em Statistica Sinica\/}~{\em 10\/}(4), 1281--1311.

\bibitem[\protect\citeauthoryear{Bengtson}{Bengtson}{2001}]{bengtson.2001}
Bengtson, V.~L. (2001).
\newblock Beyond the nuclear family: the increasing importance of
  multigenerational bonds.
\newblock {\em Journal of Marriage and the Family\/}~{\em 63\/}(1), 1--16.

\bibitem[\protect\citeauthoryear{Bland and Cook}{Bland and
  Cook}{2019}]{bland.cook.2019}
Bland, J. and A.~Cook (2019).
\newblock Random effects probit and logit: understanding predictions and
  marginal effects.
\newblock {\em Applied Economics Letters\/}~{\em 26\/}(2), 116--123.

\bibitem[\protect\citeauthoryear{Bonsang}{Bonsang}{2007}]{bonsang.2007}
Bonsang, E. (2007).
\newblock How do middle-aged children allocate time and money transfers to
  their older parents in {E}urope?
\newblock {\em Empirica\/}~{\em 34}, 171--188.

\bibitem[\protect\citeauthoryear{Bordone and de~Valk}{Bordone and
  de~Valk}{2016}]{bordone.devalk.2016}
Bordone, V. and H.~A.~G. de~Valk (2016).
\newblock Intergenerational support among migrant families in {E}urope.
\newblock {\em European Journal of Ageing\/}~{\em 13}, 259--270.

\bibitem[\protect\citeauthoryear{Brandt and Deindl}{Brandt and
  Deindl}{2013}]{brandt.deindl.2013}
Brandt, M. and C.~Deindl (2013).
\newblock Intergenerational transfers to adult children in {E}urope: do social
  policies matter?
\newblock {\em Journal of Marriage and Family\/}~{\em 75}, 235--251.

\bibitem[\protect\citeauthoryear{Chan and Ermisch}{Chan and
  Ermisch}{2015}]{chan.ermisch.2015}
Chan, T.~W. and J.~Ermisch (2015).
\newblock Residential proximity of parents and their adult offspring in the
  {U}nited {K}ingdom, 2009-10.
\newblock {\em Population Studies\/}~{\em 69\/}(3), 355--72.

\bibitem[\protect\citeauthoryear{Cheng, Birditt, Zarit, and Fingerman}{Cheng
  et~al.}{2015}]{cheng.etal.2015}
Cheng, Y.~P., K.~S. Birditt, S.~H. Zarit, and K.~L. Fingerman (2015).
\newblock Young adults' provision of support to middle-aged parents.
\newblock {\em Journals of Gerontology, Series B\/}~{\em 70\/}(3), 407--416.

\bibitem[\protect\citeauthoryear{Deindl and Brandt}{Deindl and
  Brandt}{2011}]{deindl.brandt.2011}
Deindl, C. and M.~Brandt (2011).
\newblock Financial support and practical help between older parents and their
  middle-aged children.
\newblock {\em Ageing \& Society\/}~{\em 31}, 645–62.

\bibitem[\protect\citeauthoryear{Di~Gessa, Zaninotto, and Glaser}{Di~Gessa
  et~al.}{2020}]{digessa.etal.2000}
Di~Gessa, G., P.~Zaninotto, and K.~Glaser (2020).
\newblock Looking after grandchildren.
\newblock {\em Demographic Research\/}~{\em 43}, 1545--1562.

\bibitem[\protect\citeauthoryear{Disney, Grundy, and Johnson}{Disney
  et~al.}{1997}]{disney.etal.1997}
Disney, R.~E., E.~Grundy, and P.~Johnson (1997).
\newblock {\em The Dynamics of Retirement: Analyses of the Retirement Surveys}.
\newblock Department of Social Security Research Report no. 72. London: The
  Stationery Office.

\bibitem[\protect\citeauthoryear{Dykstra}{Dykstra}{2018}]{dykstra.2018}
Dykstra, P.~A. (2018).
\newblock Cross-national differences in intergenerational family relations: the
  influence of public policy arrangements.
\newblock {\em Innovation in Aging\/}~{\em 2\/}(1), igx032.

\bibitem[\protect\citeauthoryear{Ermisch}{Ermisch}{2014}]{ermisch.2014}
Ermisch, J. (2014).
\newblock Parents' health and children's help.
\newblock {\em Advances in Life Course Research\/}~{\em 22}, 15--26.

\bibitem[\protect\citeauthoryear{Evandrou, Falkingham, Gomez-Leon, and
  Vlachantoni}{Evandrou et~al.}{2018}]{evandrou.etal.2018}
Evandrou, M., J.~Falkingham, M.~Gomez-Leon, and A.~Vlachantoni (2018).
\newblock Intergenerational flows of support between parents and adult children
  in {B}ritain.
\newblock {\em Ageing \& Society\/}~{\em 38\/}(2), 321--351.

\bibitem[\protect\citeauthoryear{Fingerman, Kim, Davis, Furstenberg, Birditt,
  and Zarit}{Fingerman et~al.}{2015}]{fingerman.etal.2015}
Fingerman, K.~L., K.~Kim, E.~M. Davis, F.~F.~J. Furstenberg, K.~S. Birditt, and
  S.~H. Zarit (2015).
\newblock ``{I}'ll give you the world'': socioeconomic differences in parental
  support of adult children.
\newblock {\em Journal of Marriage and the Family\/}~{\em 77\/}(4), 844--865.

\bibitem[\protect\citeauthoryear{Fingerman, Pillemer, Silverstein, and
  Suitor}{Fingerman et~al.}{2012}]{fingerman.etal.2012}
Fingerman, K.~L., K.~A. Pillemer, M.~Silverstein, and J.~J. Suitor (2012).
\newblock The baby boomers' intergenerational relationships.
\newblock {\em Gerontologist\/}~{\em 52\/}(2), 199--209.

\bibitem[\protect\citeauthoryear{Gardiner, Gustaffsson, Brewer, Handscomb,
  Henehan, Judge, and Rahman}{Gardiner et~al.}{2020}]{gardiner.etal.2020}
Gardiner, L., M.~Gustaffsson, M.~Brewer, K.~Handscomb, K.~Henehan, L.~Judge,
  and F.~Rahman (2020).
\newblock An intergenerational audit for the {UK}.
\newblock Report, The Resolution Foundation.

\bibitem[\protect\citeauthoryear{Gelman, Carlin, Stern, and Rubin}{Gelman
  et~al.}{2004}]{gelman.etal.2004}
Gelman, A., J.~B. Carlin, H.~S. Stern, and D.~B. Rubin (2004).
\newblock {\em Bayesian Data Analysis\/} (2nd ed.).
\newblock Boca Raton, FL: Chapman and Hall/CRC.

\bibitem[\protect\citeauthoryear{Geweke}{Geweke}{1991}]{geweke.1991}
Geweke, J. (1991).
\newblock Efficient simulation from the multivariate normal and student-t
  distributions subject to linear constraints.
\newblock In E.~M. Keramidas (Ed.), {\em Computing Science and Statistics:
  Proceedings of the 23rd Symposium on the Interface}, pp.\  571--578.
  Interface Foundation of North America, Inc., Fairfax.

\bibitem[\protect\citeauthoryear{Grootegoed and van Dijk}{Grootegoed and van
  Dijk}{2012}]{grootegoed.vandijk.2012}
Grootegoed, E. and D.~van Dijk (2012).
\newblock The return of the family? {W}elfare state retrenchment and client
  autonomy in long-term care.
\newblock {\em Journal of Social Policy\/}~{\em 41\/}(4), 677--694.

\bibitem[\protect\citeauthoryear{Grundy}{Grundy}{2005}]{grundy.2005}
Grundy, E. (2005).
\newblock Reciprocity in relationships: socio-economic and health influences on
  intergenerational exchanges between third age parents and their adult
  children in {G}reat {B}ritain.
\newblock {\em British Journal of Sociology\/}~{\em 56\/}(2), 233--255.

\bibitem[\protect\citeauthoryear{Grundy and Henretta}{Grundy and
  Henretta}{2006}]{grundy.henretta.2006}
Grundy, E. and J.~C. Henretta (2006).
\newblock Between elderly parents and adult children: a new look at the
  intergenerational care provided by the `sandwich generation'.
\newblock {\em Ageing and Society\/}~{\em 26\/}(5), 707--722.

\bibitem[\protect\citeauthoryear{Grundy and Murphy}{Grundy and
  Murphy}{2006}]{grundy.murphy.2006}
Grundy, E. and M.~Murphy (2006).
\newblock Kin availability, contact and support exchanges between adult
  children and their parents in {G}reat {B}ritain.
\newblock In F.~Ebtehaj, B.~Lindley, and M.~Richards (Eds.), {\em Kinship
  Matters}, pp.\  217--235. Oxford: Hart Publishing.

\bibitem[\protect\citeauthoryear{Grundy, Murphy, and Shelton}{Grundy
  et~al.}{1999}]{grundy.etal.1999}
Grundy, E., M.~Murphy, and N.~Shelton (1999).
\newblock Looking beyond the household: intergenerational perspectives on
  living kin and contacts with kin in {G}reat {B}ritain.
\newblock {\em Population Trends\/}~{\em 97}, 33--41.

\bibitem[\protect\citeauthoryear{Grundy and Read}{Grundy and
  Read}{2012}]{grundy.read.2012}
Grundy, E. and S.~Read (2012).
\newblock Social contacts and receipt of help among older people in {E}ngland:
  are there benefits from having more children?
\newblock {\em Journals of Gerontology: Social Sciences\/}~{\em 67\/}(6),
  742--754.

\bibitem[\protect\citeauthoryear{Grundy and Shelton}{Grundy and
  Shelton}{2001}]{grundy.shelton.2001}
Grundy, E. and N.~Shelton (2001).
\newblock Contact between adult children and their parents in {G}reat {B}ritain
  1986--1999.
\newblock {\em Environment and Planning A\/}~{\em 33}, 685--697.

\bibitem[\protect\citeauthoryear{Hajivassiliou, McFadden, and
  Ruud}{Hajivassiliou et~al.}{1996}]{hajivassiliou.etal.1996}
Hajivassiliou, V., D.~McFadden, and P.~Ruud (1996).
\newblock Simulation of multivariate normal rectangle probabilities and their
  derivatives: theoretical and computational results.
\newblock {\em Journal of Econometrics\/}~{\em 72}, 85--134.

\bibitem[\protect\citeauthoryear{Henretta, Grundy, and Harris}{Henretta
  et~al.}{2002}]{henretta.etal.2002}
Henretta, J.~C., E.~Grundy, and S.~Harris (2002).
\newblock Socio-economic and health differences in parents' provision of help
  to adult children: a {B}ritish-{USA} comparison.
\newblock {\em Ageing \& Society\/}~{\em 22}, 441--458.

\bibitem[\protect\citeauthoryear{Henretta, Van~Voorhis, and Soldo}{Henretta
  et~al.}{2018}]{henretta.etal.2018}
Henretta, J.~C., M.~F. Van~Voorhis, and B.~J. Soldo (2018).
\newblock Cohort differences in parental financial help to adult children.
\newblock {\em Demography\/}~{\em 55}, 1567--1582.

\bibitem[\protect\citeauthoryear{Heylen, Mortelmans, Hermans, and
  Boudiny}{Heylen et~al.}{2012}]{heylen.etal.2012}
Heylen, L., D.~Mortelmans, M.~Hermans, and K.~Boudiny (2012).
\newblock The intermediate effect of geographic proximity on intergenerational
  support: a comparison of {F}rance and {B}ulgaria.
\newblock {\em Demographic Research\/}~{\em 27}, 455--486.

\bibitem[\protect\citeauthoryear{Hogan, Eggebeen, and Clogg}{Hogan
  et~al.}{1993}]{hogan.etal.1993}
Hogan, D.~P., D.~J. Eggebeen, and C.~C. Clogg (1993).
\newblock The structure of intergenerational exchanges in {A}merican families.
\newblock {\em American Journal of Sociology\/}~{\em 98\/}(6), 1428--1458.

\bibitem[\protect\citeauthoryear{Huinink, Bruderl, Nauck, Walper, Castiglioni,
  and Feldhaus}{Huinink et~al.}{2011}]{huinink.etal.2011}
Huinink, J., J.~Bruderl, B.~Nauck, S.~Walper, L.~Castiglioni, and M.~Feldhaus
  (2011).
\newblock Panel analysis of intimate relationships and family dynamics
  (pairfam): conceptual framework and design.
\newblock {\em Zeitschrift f\"{u}r Familienforschung (Journal of Family
  Research)\/}~{\em 23}, 77--101.

\bibitem[\protect\citeauthoryear{Kalmijn}{Kalmijn}{2016}]{kalmijn.2016}
Kalmijn, M. (2016).
\newblock Children’s divorce and parent–child contact: a within-family
  analysis of older {E}uropean parents.
\newblock {\em The Journals of Gerontology: Series B\/}~{\em 71\/}(2),
  332–343.

\bibitem[\protect\citeauthoryear{Kalmijn}{Kalmijn}{2019}]{kalmijn.2019}
Kalmijn, M. (2019).
\newblock The effects of ageing on intergenerational support exchange: a new
  look at the hypothesis of flow reversal.
\newblock {\em European Journal of Population\/}~{\em 35}, 263--284.

\bibitem[\protect\citeauthoryear{Kalmijn and Saraceno}{Kalmijn and
  Saraceno}{2008}]{kalmijn.saraceno.2008}
Kalmijn, M. and C.~Saraceno (2008).
\newblock A comparative perspective on intergenerational support.
  {R}esponsiveness to parental needs in individualistic and familialistic
  countries.
\newblock {\em European Societies\/}~{\em 10\/}(3), 479--508.

\bibitem[\protect\citeauthoryear{Keane}{Keane}{1993}]{keane.1993}
Keane, M.~P. (1993).
\newblock Simulation estimation for panel data models with limited dependent
  variables.
\newblock In G.~S. Maddala, C.~R. Rao, and H.~D. Vinod (Eds.), {\em Handbook of
  Statistics, Vol. II}, pp.\  545--571. Elsevier Science Publishers.

\bibitem[\protect\citeauthoryear{Kim, Zarit, Eggebeen, Birditt, and
  Fingerman}{Kim et~al.}{2011}]{kim.etal.2011}
Kim, K., S.~H. Zarit, D.~J. Eggebeen, K.~S. Birditt, and K.~L. Fingerman
  (2011).
\newblock Discrepancies in reports of support exchanges between aging parents
  and their middle-aged children.
\newblock {\em Journals of Gerontology Series B\/}~{\em 66\/}(5), 527--537.

\bibitem[\protect\citeauthoryear{Lee}{Lee}{2020}]{lee.2020}
Lee, R. (2020).
\newblock Population aging and the historical development of intergenerational
  transfer systems.
\newblock {\em Genus\/}~{\em 76\/}(1).

\bibitem[\protect\citeauthoryear{Leopold and Raab}{Leopold and
  Raab}{2013}]{leopold.raab.2013}
Leopold, T. and M.~Raab (2013).
\newblock The temporal structure of intergenerational exchange: a within-family
  analysis of parent-child reciprocity.
\newblock {\em Journal of Aging Studies\/}~{\em 27\/}(3), 252--63.

\bibitem[\protect\citeauthoryear{Mudrazija}{Mudrazija}{2016}]{mudrazija.2016}
Mudrazija, S. (2016).
\newblock Public transfers and the balance of intergenerational family support
  in {E}urope.
\newblock {\em European Societies\/}~{\em 18\/}(4), 336--35.

\bibitem[\protect\citeauthoryear{Murphy}{Murphy}{2011}]{murphy.2011}
Murphy, M. (2011).
\newblock Long-term effects of the demographic transition on family and kinship
  networks.
\newblock {\em Population and Development Review\/}~{\em 37 (supplement)},
  55--80.

\bibitem[\protect\citeauthoryear{Murphy, Martikainen, and Pennec}{Murphy
  et~al.}{2006}]{murphy.etal.2006}
Murphy, M., P.~Martikainen, and S.~Pennec (2006).
\newblock Demographic change and the supply of potential family supporters in
  {B}ritain, {F}inland and {F}rance in the period 1911--2050.
\newblock {\em European Journal of Population\/}~{\em 22}, 219--240.

\bibitem[\protect\citeauthoryear{Nafilyan, Islam, Ayoubkhani, Gilles,
  Katikireddi, Mathur, Summerfield, Tingay, Asaria, John, Goldblatt, Banerjee,
  Glickman, and Khunti}{Nafilyan et~al.}{2021}]{nafilyan.etal.2021}
Nafilyan, V., N.~Islam, D.~Ayoubkhani, C.~Gilles, S.~V. Katikireddi, R.~Mathur,
  A.~Summerfield, K.~Tingay, M.~Asaria, A.~John, P.~Goldblatt, A.~Banerjee,
  M.~Glickman, and K.~Khunti (2021).
\newblock Ethnicity, household composition and {COVID-19} mortality: a national
  linked data study.
\newblock {\em Journal of the Royal Society of Medicine\/}~{\em 114\/}(4),
  182--211.

\bibitem[\protect\citeauthoryear{Nazroo}{Nazroo}{2015}]{nazroo.2015}
Nazroo, J. (2015).
\newblock Addressing inequalities in healthy life expectancy.
\newblock Report, Foresight, Government Office for Science.

\bibitem[\protect\citeauthoryear{{Race Disparity Unit}}{{Race Disparity
  Unit}}{2021}]{RDU.2021}
{Race Disparity Unit} (2021).
\newblock People in low income households. ethnicity facts and figures website.
\newblock \url{https://www.ethnicity-facts-figures.service.gov.uk}.
\newblock [posted 21 May 2021; cited 25 June 2021].

\bibitem[\protect\citeauthoryear{Rousseeuw and Molenberghs}{Rousseeuw and
  Molenberghs}{1994}]{rousseeuw.molenberghs.1994}
Rousseeuw, P.~J. and G.~Molenberghs (1994).
\newblock The shape of correlation matrices.
\newblock {\em The American Statistician\/}~{\em 48\/}(4), 276--279.

\bibitem[\protect\citeauthoryear{Schans and Komter}{Schans and
  Komter}{2010}]{schans.komter.2010}
Schans, D. and A.~Komter (2010).
\newblock Ethnic differences in intergenerational solidarity in the
  {N}etherlands.
\newblock {\em Journal of Aging Studies\/}~{\em 24}, 194--203.

\bibitem[\protect\citeauthoryear{Schoon}{Schoon}{2015}]{schoon.2015}
Schoon, I. (2015).
\newblock Diverse pathways: rethinking the transition to adulthood.
\newblock In P.~R. Amato, A.~Booth, S.~M. McHale, and J.~Van~Hook (Eds.), {\em
  Families in an Era of Increasing Inequality: Diverging Destinies}, pp.\
  115--36. Springer International Publishing.

\bibitem[\protect\citeauthoryear{Seltzer and Bianchi}{Seltzer and
  Bianchi}{2013}]{seltzer.bianchi.2013}
Seltzer, J.~A. and S.~M. Bianchi (2013).
\newblock Demographic change and parent-child relationships in adulthood.
\newblock {\em Annual Review of Sociology\/}~{\em 39}, 275--290.

\bibitem[\protect\citeauthoryear{Shapiro}{Shapiro}{2004}]{shapiro.2004}
Shapiro, A. (2004).
\newblock Revisiting the generation gap: exploring the relationships of
  parent/adult-child dyads.
\newblock {\em International Journal of Aging and Human Development\/}~{\em
  58}, 127--146.

\bibitem[\protect\citeauthoryear{Silverstein and Bengtson}{Silverstein and
  Bengtson}{1997}]{silverstein.bengtson.1997}
Silverstein, M. and V.~Bengtson (1997).
\newblock Intergenerational solidarity and the structure of adult
  child–parent relationships in {A}merican families.
\newblock {\em American Journal of Sociology\/}~{\em 103\/}(2), 429--460.

\bibitem[\protect\citeauthoryear{Silverstein, Conroy, Wang, Giarrusso, and
  Bengtson}{Silverstein et~al.}{2002}]{silverstein.etal.2002}
Silverstein, M., S.~J. Conroy, H.~Wang, R.~Giarrusso, and V.~L. Bengtson
  (2002).
\newblock Reciprocity in parent–child relations over the adult life course.
\newblock {\em Journal of Gerontology: Social Sciences\/}~{\em 57B\/}(1),
  S3--S13.

\bibitem[\protect\citeauthoryear{Snijders and Bosker}{Snijders and
  Bosker}{2012}]{snijders.bosker.2012}
Snijders, T. and R.~Bosker (2012).
\newblock {\em Multilevel Analysis: An Introduction to Basic and Advanced
  Multilevel Modeling\/} (2nd ed.).
\newblock London: Sage Publishers.

\bibitem[\protect\citeauthoryear{Steele, Clarke, and Kuha}{Steele
  et~al.}{2019}]{steele.etal.2019}
Steele, F., P.~S. Clarke, and J.~Kuha (2019).
\newblock Modeling within-household associations in household panel studies.
\newblock {\em Annals of Applied Statistics\/}~{\em 13\/}(1), 367--392.

\bibitem[\protect\citeauthoryear{Steele and Grundy}{Steele and
  Grundy}{2021}]{steele.grundy.2021}
Steele, F. and E.~Grundy (2021).
\newblock Random effects dynamic panel models for unequally spaced multivariate
  categorical repeated measures: an application to child–parent exchanges of
  support.
\newblock {\em Journal of the Royal Statistical Society C (Applied
  Statistics)\/}~{\em 70}, 3--23.

\bibitem[\protect\citeauthoryear{Steinbach}{Steinbach}{2012}]{steinback.2012}
Steinbach, A. (2012).
\newblock Intergenerational relations across the life course.
\newblock {\em Advances in Life Course Research\/}~{\em 17\/}(3), 93--99.

\bibitem[\protect\citeauthoryear{Suitor, Gilligan, Pillemer, Fingerman, Kim,
  Silverstein, and Bengtson}{Suitor et~al.}{2017}]{suitor.etal.2017}
Suitor, J.~J., M.~Gilligan, K.~Pillemer, K.~L. Fingerman, K.~Kim,
  M.~Silverstein, and V.~L. Bengtson (2017).
\newblock Applying within-family differences approaches to enhance
  understanding of the complexity of intergenerational relations.
\newblock {\em The Journals of Gerontology: Series B\/}~{\em 73\/}(1), 40--53.

\bibitem[\protect\citeauthoryear{Swartz}{Swartz}{2009}]{swartz.2009}
Swartz, T.~T. (2009).
\newblock Intergenerational family relations in adulthood: patterns,
  variations, and implications in the contemporary {U}nited {S}tates.
\newblock {\em Annual Review of Sociology\/}~{\em 35}, 191–212.

\bibitem[\protect\citeauthoryear{{University of Essex, ISER}}{{University of
  Essex, ISER}}{2020}]{UKHLS20}
{University of Essex, ISER} (2020).
\newblock {\em {\rm Understanding Society: Waves 1-10, 2009-2019 and Harmonised
  BHPS:\ Waves 1-18, 1991-2009. [data collection]}\/} (13th ed.).
\newblock University of Essex, Institute for Social and Economic Research. UK
  Data Service. SN: 6614, \url{http://doi.org/10.5255/UKDA-SN-6614-13}.

\bibitem[\protect\citeauthoryear{Vlachantoni, Evandrou, Falkingham, and
  Gomez-Leon}{Vlachantoni et~al.}{2020}]{vlachantoni.etal.2020}
Vlachantoni, A., M.~Evandrou, J.~Falkingham, and M.~Gomez-Leon (2020).
\newblock Caught in the middle in mid-life: provision of care across multiple
  generations.
\newblock {\em Ageing \& Society\/}~{\em 40\/}(7), 1490--1510.

\bibitem[\protect\citeauthoryear{Wiemers and Bianchi}{Wiemers and
  Bianchi}{2015}]{wiemers.bianchi.2015}
Wiemers, E.~E. and S.~M. Bianchi (2015).
\newblock Competing demands from aging parents and adult children in two
  cohorts of {A}merican women.
\newblock {\em Population and Development Review\/}~{\em 41\/}(1), 127--146.

\bibitem[\protect\citeauthoryear{Willetts}{Willetts}{2019}]{willetts.2019}
Willetts, D. (2019).
\newblock {\em The Pinch: How the Baby Boomers Took Their Children's Future
  –- And Why They Should Give It Back\/} (2nd ed.).
\newblock London: Atlantic Books.

\bibitem[\protect\citeauthoryear{Willis}{Willis}{2012}]{willis.2012}
Willis, R. (2012).
\newblock Individualism, collectivism and ethnic identity: cultural assumptions
  in accounting for caregiving behaviour in {B}ritain.
\newblock {\em Journal of Cross-Cultural Gerontology\/}~{\em 27}, 201--216.

\bibitem[\protect\citeauthoryear{Wolf}{Wolf}{1994}]{wolf.1994}
Wolf, D.~A. (1994).
\newblock The elderly and their kin: patterns of availability and access.
\newblock In L.~G. Martin and W.~H. Preston (Eds.), {\em Demography of Aging},
  pp.\  146--194. National Research Council (US) Committee on Population;
  Washington (DC): National Academies Press (US).

\bibitem[\protect\citeauthoryear{Zhang and Steele}{Zhang and
  Steele}{2022}]{mvreprobit}
Zhang, S. and F.~Steele (2022).
\newblock \texttt{mvreprobit}: efficient {G}ibbs sampling procedures for
  multivariate random effect probit model estimation.
\newblock \url{https://github.com/slzhang-fd/mvreprobit}.
\newblock [{R} package, version 0.1.0].

\bibitem[\protect\citeauthoryear{Zigante, Fernandez, and Mazzotta}{Zigante
  et~al.}{2021}]{zigante.etal.2021}
Zigante, V., J.~L. Fernandez, and F.~Mazzotta (2021).
\newblock Changes in the balance between formal and informal care supply in
  {E}ngland between 2001 and 2011: evidence from census data.
\newblock {\em Health Economics, Policy and Law\/}~{\em 16\/}(2), 232--49.

\end{thebibliography}
\bibliographystyle{chicago}

\end{document}